\documentclass{aa}
\usepackage{graphicx}
\usepackage{txfonts}
\begin{document}
   \title{HH135/HH136 - a luminous H$_2$ outflow towards a 
high-mass protostar
}

%   \subtitle{}

   \author{R. Gredel
          \inst{1}
          }

   \offprints{R. Gredel}

   \institute{Max-Planck Institut f\"ur Astronomie, 
	      K\"onigstuhl 17, D-69117 Heidelberg, Germany\\
              \email{gredel@mpia.de}
             \thanks{
Based on observations collected at the 
European Southern Observatory, La Silla, Chile. Figures .1--.4 and 
Tables .1--.9 are only
available in electronic form via http://www.edpsciences.org
}
             }

   \date{Received; accepted}

\abstract
%
%1. context
%
{Molecular hydrogen observations towards Herbig-Haro objects provide 
the possibility of studying physical processes related to star formation.}
%
%2. aims
%
{Observations towards the luminous IRAS source IRAS11101-5928 and 
the associated Herbig-Haro objects HH135/HH136 are obtained  
to understand whether high-mass stars form via the same 
physical processes as their low-mass counterparts. 
}
%
%3. methods
%
{Near-infrared imaging and spectroscopy are used to infer H$_2$ excitation 
characteristics. A theoretical H$_2$ spectrum is constructed from a
thermal ro-vibrational population distribution and compared to the 
observations.
}
%
%4. results
%
{The observations reveal the presence of 
a well-collimated, parsec-sized  H$_2$ outflow
with a total H$_2$ luminosity of about $2L_\odot$. 
The bulk of the molecular gas is characterized 
by a ro-vibrational excitation temperature of $2000\pm200$~K. A 
small fraction (0.3\%) of the molecular gas is very hot, with
excitation temperatures around 5500~K.
The molecular emission is associated with strong
[FeII] emission.
The H$_2$ and [FeII] emission characteristics 
indicate the presence of fast, 
dissociative J-shocks at speeds of $v_\mathrm{s} \approx$ 100 km s$^{-1}$.
Electron densities of $n_\mathrm{e}$ = 3500-4000 cm$^{-3}$ are inferred
from the [FeII] line ratios.}
%
%5. conclusions
%
{The large H$_2$ luminosity combined
with the very 
large source luminosity suggests that the high-mass protostar
that powers the HH135/HH136 flow forms via accretion, but with a
significantly increased accretion rate compared to that
of low-mass protostars.
}
\keywords{
ISM:individual objects: HH135/HH136
- ISM: Herbig-Haro objects
- ISM: jets and outflows
}

   \maketitle
%
%________________________________________________________________

\section{Introduction}

Near-infrared studies of molecular hydrogen emission in 
Herbig-Haro (HH) objects provide a powerful tool to gain insight in
the physical processes that occur during the early phases of 
low-mass star formation. The
total H$_2$ luminosities are proportional to the
accretion rates in the early phases of the protostellar 
evolution (Stanke \cite{stanke}; Froebrich et al. \cite{froebrich}),
and the H$_2$ ro-vibrational population distribution allows us to 
determine the physical scenarios that are at work in the  
supersonic jets that form during the accretion phase 
(McCoey \cite{mccoey}, and references therein).  A complementary
tool to study such regions is available via the near-infrared emission 
lines of [FeII]
and other atomic emission lines,
that allow us to measure electron densities 
in the shocked material (e.g., Nisini et al. \cite{nisini}). 

In a number of HH objects, and most 
prominently in HH objects that show pronounced [FeII] emission, 
a temperature stratification is inferred from the H$_2$ observations, 
where part of the molecular gas reach
temperatures above 5000~K (Giannini et al. \cite{giannini04}). 
The combined H$_2$ and [FeII] emission has been explained in
terms of fast J-type shocks, where the 
[FeII] emission is produced in dissociative parts of the shocks, 
and where H$_2$ arises in the slower, non-dissociative regions
(e.g., Gredel \cite{gredel94}). Detailed model calculations 
of J-type shocks with magnetic precursors confirm such a model, and
produce population distributions 
where the rotational excitation temperatures increase with increasing
vibrational state of H$_2$ (e.g., Flower et al. \cite{flower}). 
C-type shocks at largely different physical conditions produce similar
H$_2$ population distributions, however, and the H$_2$ population distribution
alone does not allow us to distinguish between both
scenarios (Flower et al. \cite{flower}). The presence of [FeII] emission
is generally interpreted in terms of dissociative J-type shocks, 
although it is conceivable that C-type shocks produce [FeII] as well
(Le Bourlot et al. \cite{lebourlot}). 

Near-infrared studies of outflows from intermediate- and high-mass star
forming regions are rare, either because such outflows are not very
frequent {\tt per se} or because high-mass star-forming regions are
deeply embedded in general. 
Some outflows that are observed from high-luminous sources such as 
Orion ($10^5 L_\odot$) lack the high degree of collimation that
is typical for outflows from low-mass star-forming regions. 
Other flows that emerge from luminous IRAS sources, such as
HH80/81 ($2 10^4 L_\odot$, 
Mart\'i, Rodr\'iguez, \& Reipurth \cite{marti}), IRAS 16547-4247
($6 10^4 L_\odot$, Brooks et al. \cite{brooks03}), or IRAS 18151-1208
($2 10^4 L_\odot$, Davis et al. \cite{davis04}), are highly collimated. 
It is not clear whether intermediate- and high-mass stars form in
a similar way to low-mass stars, but with enhanced accretion rates,
or whether different processes, such as the merging of low-mass 
protostars, are at work. Enlarging the sample
of near-infrared observations of regions of intermediate- and high-mass
star formation and comparing the general properties of their H$_2$
and [FeII] emission with those of low-mass star-forming regions is
therefore desirable.  
 
In the following, a study of the molecular outflow 
\object{HH135}/\object{HH136}, 
which is powered by the cold and very luminous ($10^4 L_\odot$) 
IRAS source \object{IRAS11101-5928}, 
is presented.  The pair of Herbig-Haro objects HH135/HH136 was discovered by
Ogura \& Walsh (\cite{ogura}) in an objective prism survey and
is located in the bright rimmed cloud No. 64 of Sugitani \& Ogura 
(\cite{sugitani}) in the Eastern Carina region.
The HII region, also known as Gum 36, is believed
to be excited by the open cluster Stock~13, for which photometric distances
of 2.7 kpc are available (Steppe \cite{steppe}). 
The general morphology of HH135/HH136
indicates that the two objects are formed at the
opposite flow directions of a bipolar flow, which is driven by
IRAS11101-5829.  The velocity field
of the HH135/HH136 region is complex, however, and the fact that
the main part of the emission from both HH135 and HH136 is
blue shifted led Ogura \& 
Walsh (\cite{ogura}) to conclude that HH135 and HH136 form two 
different, independent flows.  Near-infrared JHK polarimetric
observations of the associated reflection nebula 
carried out by Tamura et al. (\cite{tamura}) showed that 
IRAS11101--5829 is the only illuminating source of the
nebula that is associated with HH135/HH136. 
A more recent millimetre study by
Ogura et al. (\cite{ogura98}) presented  a model that explains
the observed velocity features and where HH135 and
HH136 form part of a single, bipolar flow driven
by IRAS11101-5829. This view is supported by the very recent polarimetric 
observations by Chrysostomou et al.  (\cite{chrysostomou}),
who proposed that a strong helical magnetic field threading though HH135/HH136
maintains the strong collimation of the flow.  

   \begin{figure}
   \centering
   \includegraphics[angle=-90,width=12cm]{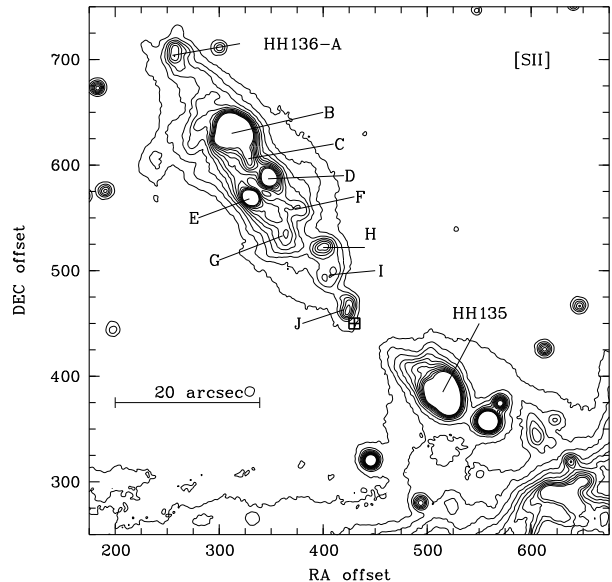}
   \caption{[SII] image of HH135/HH136 and part of the G~36 region in Carina.  
   Optical knots in HH136 are labeled following the notation by 
   Ogura \& Walsh (\cite{ogura}). 
{The crossed square south of knot J marks the position of 
the IRAS source IRAS11101-5829.
}
North is up and east is left.  
}
              \label{img_SII}
    \end{figure}

   \begin{figure}
   \centering
   \includegraphics[angle=-90,width=12cm]{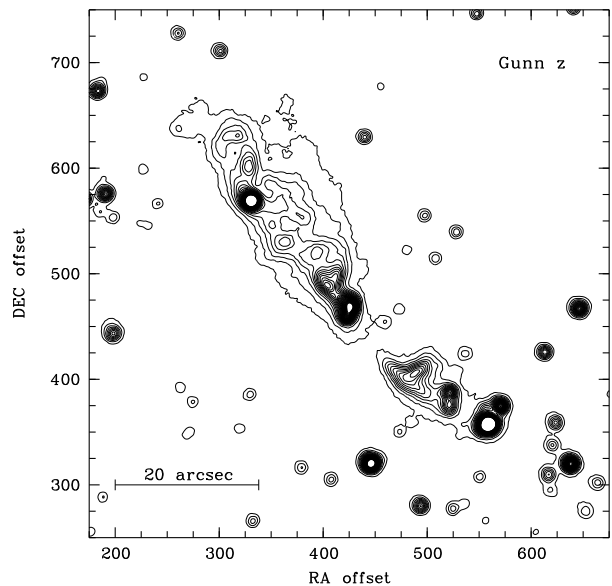}
   \caption{Gunn z image of HH135/HH136.  
}
              \label{img_gunnz}
    \end{figure}

The purpose of this paper is to investigate the molecular hydrogen emission 
of HH135/HH136, and to study how the emission characteristics differ
from those of low-mass star-forming
regions. The observations are presented in Sect.~\ref{observations}, 
which also includes a more detailed discussion of the data reduction
method and errors. The imaging results are given in 
Sect.~\ref{imaging} and the spectroscopy  in Sect.~\ref{spectroscopy}. 
Section~\ref{sumspectra}
contains an analysis of the global properties of the molecular hydrogen
emission, and Sect.~\ref{FeII} summarises the results of an analysis derived from
the [FeII] lines, which are observed towards HH135/HH136. 
The main conclusions of this study are summarized in Sect.~\ref{conclusions}.

%__________________________________________________________________

\section{Observations and reduction}
\label{observations}

Near-infrared imaging and spectroscopy of the HH135/HH136 complex was
carried out in March 1999 and February 2004 at the La Silla Observatory,
using SOFI at the New Technology Telescope (NTT). 
Images were obtained at both epochs in three narrow-band filters 
NB164, NB212, and
NB228, using a pixel scale of $0\farcs144$/px.
 Central wavelengths are 1.644 $\mu$m, 2.124 $\mu$m, and 
2.280 $\mu$m, respectively, and widths are 0.025 $\mu$m, 0.028 $\mu$m,
and 0.030 $\mu$m,  respectively, for the NB164, NB212, and NB228 filters,
according to the information given in the ESO web pages. 
Dithered images were used to construct background
subtracted images, including corrections for sky, dark-current, and bias. 
This method assumes  that the background is
constant and does not vary across an image, which is not the case in the
H-band. All images were re-centered to a
common reference, but no effort was made to correct for field distortion
across the image. 

In March 1999, spectroscopic observations were obtained
along three slit positions 
using the red grism GR in order 1, covering the 1.52--2.52 $\mu$
wavelength region. The slit was set to a width of $0\farcs6$, and measurements of 
the width of individual spectral lines showed that an effective spectral resolution
of R = $\lambda / \Delta\lambda$ = 550 was achieved. Spectroscopic observations
were also carried out in February 2004, using the blue grism GB in order 1 
and a slit of $0\farcs6$,
covering the 0.94--1.60 $\mu$m wavelength region at a spectral resolution of 
R=850. Additional spectra were gathered using
the HR grism in orders 3 and 4, covering the 2.0--2.3 $\mu$m and 1.5--1.8 $\mu$m
wavelength regions, respectively. The K- and H-band
observations obtained with the HR grism were carried out with a slit width
of $0\farcs6$ as well. 
Spectral resolutions, as obtained from the width of the
emission lines, were R=1066 and R=1600, respectively. 
Explicit {\tt sky} frames were obtained by moving the object along the slit.
This is possible because the infrared emission of HH135/HH136 is extended 
over about 1 arcmin, compared to the total slit length of 5 arcmin. 

The spectroscopic observations were reduced using the MIDAS {\tt long}
package.  The long-slit spectra suffer from a pronounced curvature of the
spectral lines perpendicular to the dispersion direction, and the distorted
2-D frames were rectified by use of a second-order polynomial. 
Pixel units along the dispersion direction were then 
transformed into a linear wavelength scale by the use of Xenon-Argon 
calibration frames. A first-order background subtraction
was obtained by building ({\tt object - sky}) difference frames.
Short-term temporal and spatial variations at scales comparable to
the length of the slit result in residuals in the background in a large
number of the difference frames. A second-order sky subtraction was
achieved via the standard technique of extracting
the positive and negative spectra from the difference frames and adding
them in the correct sense. 
Extraction windows to obtain the 1-D spectra from the
2-D difference frames were typically 12 pixels along the
slit. On the spatial scale of $0\farcs29$/px, the
extraction windows correspond to solid angles $\Omega$ of
about 2 arcsec$^2$ on the sky or $\Omega = 5 10^{-11}$ sr.  
The atmospheric transmission
was determined via observations of standard stars at slit widths
of $0\farcs6$. Stellar fluxes were measured with a second integration,
 but slits opened to $2''$ to minimise slit losses. 
During the observations, the seeing varied from $0\farcs5$ to
$1\farcs5$. In phases of poor seeing, the flux conversion factors 
determined from the stellar observations varied by up to 30\%. 
{
The highly peaked nature of the line emission in some of
the knots, combined
with positional errors in the slit position for the various grism
settings and the narrow slit of $0\farcs6$ adopted here, causes
differences in the reported fluxes towards some of the knots that
reach, in the worst cases, a factor of about three (cf. Tables~
\ref{all_H} and \ref{all_GR}). This is well illustrated 
in Fig.~\ref{img_bowNE}, which shows that an error of less than
$0\farcs5$ in the position of the slit would cause the loss of
most of the flux in knot B1. 
The consequences of these kinds of flux uncertainties
for the analysis of the H$_2$ and atomic emission lines in 
HH135/HH136 are discussed in detail in Sects.~\ref{spectroscopy}
and \ref{FeII}.
}
The tables of Pickles (\cite{pickles}) were used to convert the 
counts of the standard stars to flux units. 

In addition to the near-infrared observations, optical images 
in the [SII] ($\lambda_\mathrm{center}$ = 673nm, FWHM = 6nm)
and Gunn z($\lambda_\mathrm{cut-on}$ = 840 nm)
were obtained in April 1993, using EFOSC2 at the 2.2m telescope 
and CCD\#19. Integration times were 300~sec in [SII] and 2100~sec
in Gunn z. 

\section{Results and discussion}
\label{results}

The narrow-band [SII] image of the HH135/HH136 region 
and the Gunn z image are shown in Figs.~\ref{img_SII} and \ref{img_gunnz}. 
The observations were obtained during 
an atmospheric seeing of about $1\farcs3$. The image was re-binned to a
pixel scale of $0\farcs144$ to allow for a comparison with the 
near-infrared images described below. The optical knots in
the [SII]-image are labeled following the notation of Ogura \& Walsh 
(\cite{ogura}). A continuum subtracted image 
constructed by use of the Gunn z image shows 
that the optical knots HH136-E (IRS9 in the notation
of Tamura et al. \cite{tamura}), HH136-I, and HH136-J are continuum
sources with weak if any [SII] emission. Strong but featureless 
[SII]-emission  occurs towards HH135. The [SII] emission at the southern 
boundary of the imaged region traces the bright rim of Sugitani~64. 

   \begin{figure}
   \centering
   \includegraphics[angle=-90,width=12cm]{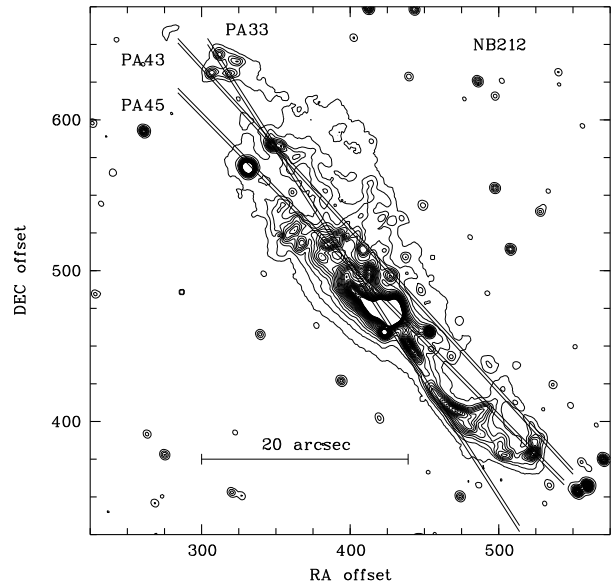}
   \caption{Narrow-band image of the HH135/HH136 region obtained in the
NB212 filter, which is centered on the (1,0) S(1) H$_2$ emission line 
near 2.12 $\mu$m. The three double lines indicate slit locations
and slit widths
for the February 2004 spectroscopic observations, with position angles of
PA = 33$^\circ$, 43$^\circ$, and 45$^\circ$. North is up and east is left.
}
              \label{img_212}
    \end{figure}

   \begin{figure}
   \centering
   \includegraphics[angle=-90,width=12cm]{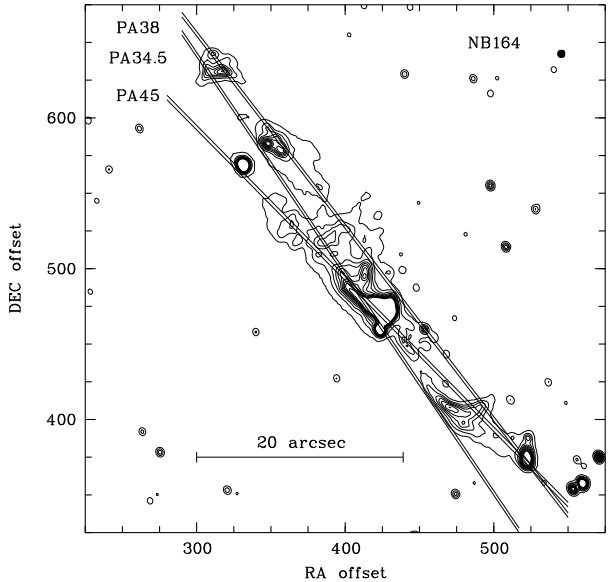}
   \caption{Narrow-band image of the HH135/HH136 region obtained in the
NB164 filter, which is centered on the wavelength of the [FeII] 
$a^4D_{7/2} - a^4F_{9/2}$ emission line near 1.644 $\mu$m. 
The three double lines indicate slit locations and slit widths 
for the March 1999 spectroscopic observations, with position angles of 
PA = 34.$^\circ$5, 38$^\circ$, and 45$^\circ$. North is up and east is
left.  
}
              \label{img_164}
    \end{figure}

   \begin{figure}
   \centering
   \includegraphics[angle=-90,width=12cm]{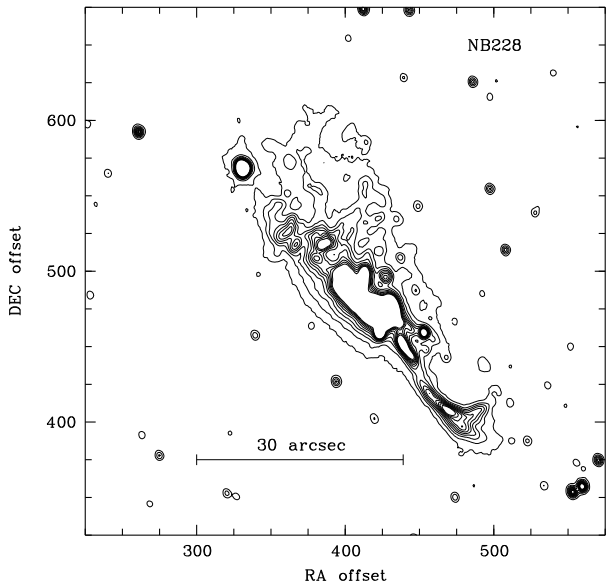}
   \caption{Narrow-band image of the HH135/HH136 region obtained in the
NB228 filter. North is up and east is left.
}
              \label{img_228}
    \end{figure}

\subsection{Near-infrared imaging of HH135/HH136}
\label{imaging}

\begin{figure}
   \centering
\includegraphics[angle=-90,width=12cm]{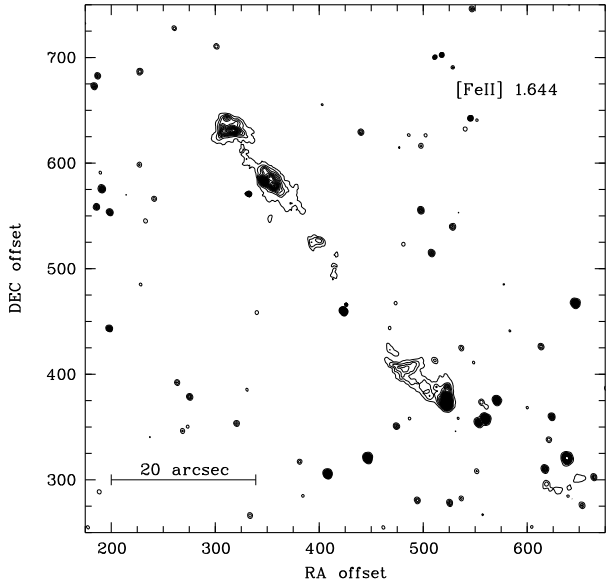}
      \caption{Continuum-subtracted image showing the [FeII] 1.64 $\mu$m
emission towards HH135/HH136.
              }
         \label{img_FeII}
   \end{figure}

\begin{figure}
   \centering
\includegraphics[angle=-90,width=12cm]{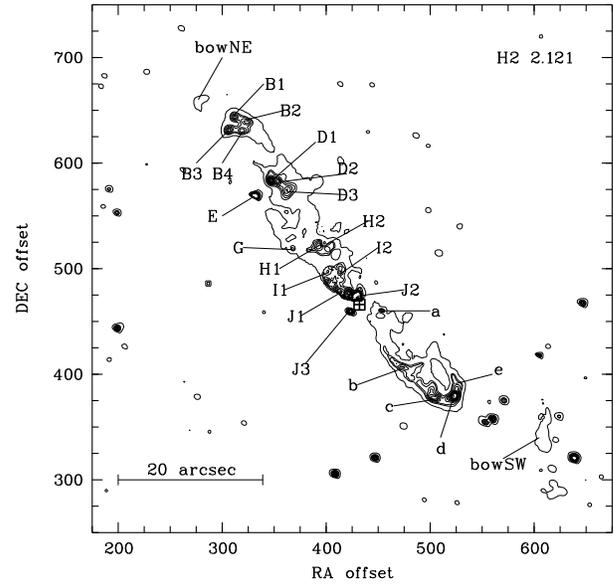}
      \caption{Continuum-subtracted image showing the H$_2$ 2.122 $\mu$m
emission towards HH135/HH136. The strongest knots are labeled. 
              }
         \label{img_H2}
   \end{figure}

Narrow-band images obtained in the NB212, NB164, and NB228 filters
are shown in Figs.~\ref{img_212}, \ref{img_164}, and \ref{img_228}, 
respectively.  
The slit locations shown in Fig.~\ref{img_212} are those adopted
during the February 2004 observations and correspond to
position angles of $33^\circ$, $43^\circ$, and $45^\circ$.
Figure~\ref{img_164} includes
the slit locations for the March 1999 spectroscopic observations,
at position angles of $34^\circ.5$, $38^\circ$, and $45^\circ$. 
Observations at a position angle of $34^\circ.5$ were repeated
twice. The NB228 passband does not cover any strong emission lines 
and was used to construct the continuum-subtracted images in the 
H$_2$ 2.122 $\mu$m and [FeII] 1.644 $\mu$m lines. 
A scaling factor of 0.933 that
corresponds to the ratio of the filter widths was applied to 
the flux in NB228 to obtain the H$_2$ image shown in Fig.~\ref{img_H2}.
In a similar way, the continuum-subtracted [FeII] image shown 
in Fig.~\ref{img_FeII} was constructed using a scaling factor of 0.3. 

The overall morphology of the H$_2$ emission in the HH135/HH136 
region is that of a highly collimated bipolar flow with various H$_2$ 
emission knots along the flow direction. 
The labeling of the infrared H$_2$ knots of Fig.~\ref{img_H2}
follows the notation of 
the optical knots in the [SII] image. This is possible because in general,
the overall morphology of the [SII] and the H$_2$ emission is very similar. 
Exceptions occur for the optical knots HH136-A, C, and F, which do not have 
near-infrared counterparts in H$_2$ or [FeII], and which do not 
show significant infrared continuum emission either. 
The optical knot HH136-B splits up into four point-like objects, B1-B4.
At the NE prolongation of knots B1 and B2,
the diffuse H$_2$ emission labeled bow-NE has a 
pronounced bow-shaped morphology that suggests a B2-bowNE flow
direction.  Knots D1-D3 (optical knot HH136-D)
show strong H$_2$ emission with weak underlying continua. 
Knot E is a strong continuum source (IRS-9 in the notation of Tamura et al.
(\cite{tamura})) without pronounced H$_2$ emission. The region between
infrared knots G-J2 contains a large number of infrared continuum sources,
and very strong infrared continuum emission occurs between knots H1-J2
(cf. Fig.~\ref{img_228}). In the NB228 continuum image shown in 
Fig.~\ref{img_228}, the presence of a large number (30-40) of
point-like continuum sources is discerned, in agreement with the finding of 
Tamura et al. (\cite{tamura}) based on their K$_\mathrm s$ image.
The slit traces shown in Figs.~\ref{trace_grf} and \ref{trace_ks} 
show that most of the continuum sources that are located 
between knots I1-J3 show H$_2$ emission  as well.  Knots J1-J3 
are very strong continuum sources. Because of the very strong
continuum emission between I1-J3,
a proper continuum-subtracted H$_2$ image for that region could not be
constructed. It is noted that between knots J2 and d, a remarkably
circular shape of very faint H$_2$ emission is discerned in a
false-color image, with knot b at the southeastern and knot a at the 
northeastern extremes of this 'cavity'. 

Five infrared knots here labeled a-e are identified in HH135. 
The H$_2$ emission of HH135 has a morphology of a bow with an apex near 
knot d and wings formed by knots b, c, and e. At the SW prolongation
of the HH136-HH135 complex, the diffuse H$_2$ emission labeled bow-SW
has a pronounced bowshape.  Faint H$_2$ emission also occurs in the 
SW extreme of HH135. The emission has a pronounced bow morphology as well. 
The H$_2$ continuum-subtracted image is reproduced in Fig.~\ref{img_bowSW}.
The bow morphology, which cannot be demonstrated too convincingly in the 
contour plot shown here, is more apparent in a false-color H$_2$ image. 
The whole extent of the collimated molecular hydrogen flow 
is 80", which corresponds to about 1 parsec at a distance of 2.7 kpc. 

The morphology of the [FeII] emission is given in Fig.~\ref{img_FeII}. 
The [FeII] emission is, in general, very strong, and it is 
associated with molecular hydrogen emission,
Exceptions occur for bow-NE and bow-SW, which show faint [FeII] emission only.
The very strong [FeII] emission near knot d
is displaced by a few arcsec towards the SW relative to its H$_2$ 
counterpart. The displacement between the [FeII] and H$_2$ emission
provides evidence for the presence of fast, 
dissociative J-type shocks with a NE-SW flow direction, where the [FeII]
emission arises in the fast, dissociative parts of the shocks, such as the apex
of a bow shock, and where H$_2$ arises in slower regions or in oblique shocks,
such as those in the wings of a bowshock. 

\begin{figure}
   \centering
\includegraphics[angle=-90,width=12cm]{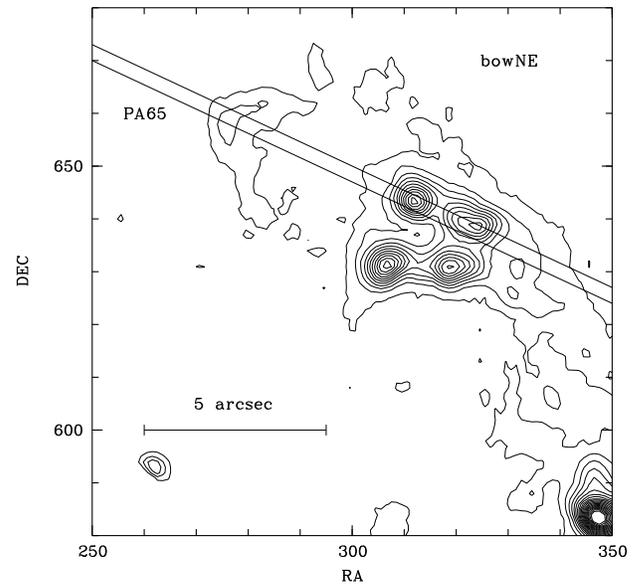}
      \caption{Morphology of the molecular hydrogen emission at the
NE extreme of the HH135/HH136 flow. The double lines indicate the position 
of a slit and its width placed at a position angle of PA=65$^\circ$. 
              }
         \label{img_bowNE}
   \end{figure}

\begin{figure}
   \centering
\includegraphics[angle=-90,width=12cm]{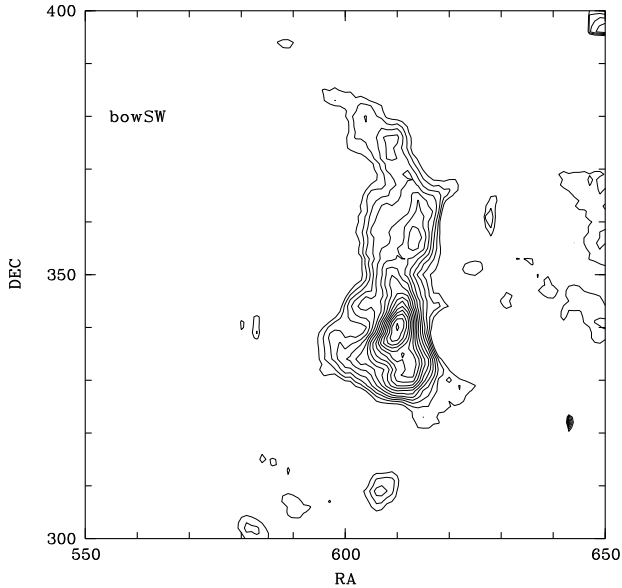}
      \caption{Morphology of the molecular hydrogen emission at the
SW extreme of the HH135/HH136 flow. 
              }
         \label{img_bowSW}
   \end{figure}

The NE-extreme of the HH135/HH136 flow is shown in Fig.~\ref{img_bowNE}, 
together with the location of a slit at position angle PA=65$^\circ$,
which was used in February 2004 to obtain spectra of objects bowNE and knots
B1 and B2. The image shows the continuum-subtracted H$_2$ emission. 
Object bowNE has a pronounced bow shape with knots B1 and B2
located along a possible SW-NE flow direction. The alignment of bowNE-B2
differs by some 20$^\circ$ from the large-scale alignment of the HH135/HH136 complex,
which suggests the presence of a second flow, that is not powered
by IRAS~11101-5829.

The images shown in Figs.~\ref{img_164}--\ref{img_H2} 
were obtained in February 2004 at a seeing of $0\farcs55$.
Seeing during the March 1999 imaging campaign was significantly worse and generally 
above 1 arcsec. A comparison of the images obtained at both epochs 
does not show any spatial displacement of the various knots, which supports
the assumption that HH135/HH136 is located at the large distance of 2.7 kpc.
Thus, no effort was made
to obtain proper motions and displacement vectors from both epochs.

\clearpage

\subsection{Molecular hydrogen excitation across the HH135/HH136 flow}
\label{spectroscopy}

During the March 1999 and February 2004 observing campaigns, spectra
of the HH135/HH136 complex were obtained using a total of 7 slit positions
at various position angles across the region.
The relatively large number of slit positions was adopted to
investigate possible changes in the H$_2$ excitation across the HH135/HH136
flow. 
The variation of the flux in the (1,0) S(1) line of H$_2$ along the various
slit positions is shown in Figs.~\ref{trace_grf}--\ref{trace_ks}. The
individual emission knots, as identified in Fig.~\ref{img_H2}, are marked 
in these slit traces. 
The bold lines represent the flux variation in 
the continuum along the slit. It can be seen that many of the H$_2$ emission 
line knots coincide with faint, underlying continuum sources.
The H- and K$_s$-band spectra are shown in Figs.~\ref{spectra_H} and
 \ref{spectra_ks}.  Note the very strong continuum emission that is 
associated with knots H1 and I2.  

Tables~\ref{all_GB} - \ref{all_GR} contain the unreddened 
line fluxes measured towards the various knots. 
From the analysis of the [FeII] emission lines, a 
visual extinction of $A_\mathrm{V} = 2.7$ mag is inferred (cf. Sect.~\ref{FeII})
and was used to obtain the population densities $N\mathrm{(v'J')}$ 
in the ro-vibrational levels
$\mathrm{v'J'}$. $A(\mathrm{v'J'v''J''})$ 
and $\bar{\nu}$ are the transition probabilities and wave numbers, respectively. 
The visual extinction obtained from the [FeII] lines is in excellent agreement 
with the values of
$A_\mathrm{V}= 2.8 - 3.1$ mag obtained by Ogura \& Walsh (\cite{ogura})
from their optical spectroscopy.
The population densities were used to
construct excitation diagrams for the various knots, where values of
$ln (N\mathrm{(v'J')/g}$ are plotted versus excitation energies 
$E\mathrm{(v'J')}$ of the
ro-vibrational levels. Statistical weights are 
$\mathrm{g = g_s (2J'+1)}$,  where the
nuclear-spin statistical weights $\mathrm{g_s}$ 
are 1 and 3 for even and odd rotational
levels J', respectively. 
The above analysis implies that the H$_2$ ortho/para ratio is 3, which in J-type shocks
is attained at shock velocities above $v_\mathrm{s} 
\approx 15$ km s$^{-1}$, or $v_\mathrm{s} > 40$ km s$^{-1}$
for C-type shocks (Wilgenbus et al. \cite{wilgenbus}). 

\begin{figure*}
\sidecaption
   \includegraphics[width=15cm]{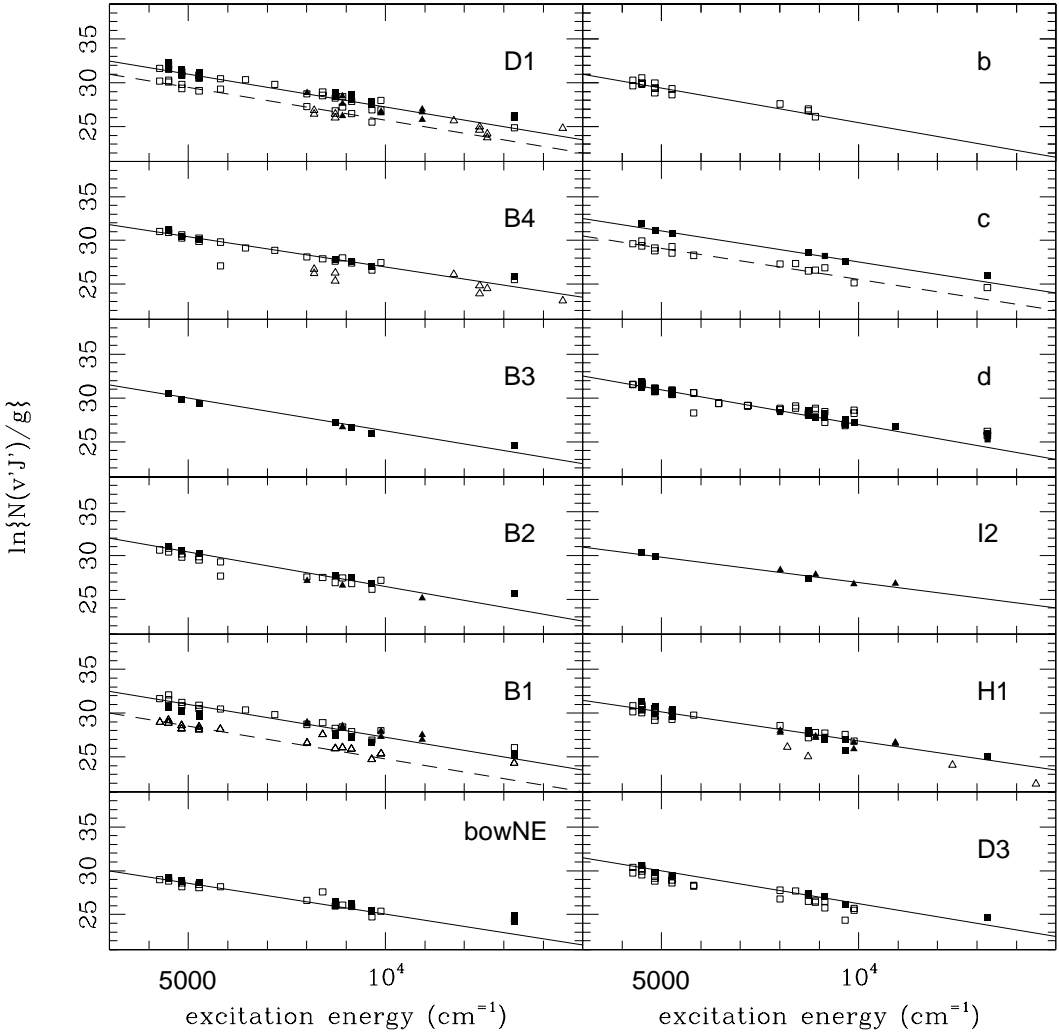}
   \caption{Excitation diagrams constructed for individual knots in HH135/HH136
   with values of $ln{(N(v'J')/g}$ plotted versus excitation energies $E(v'J')$. 
   Open triangles and open squares correspond to data inferred from the blue and 
   red grisms GB and GR, respectively, while filled triangles and squares 
   represent the data inferred from the H- and K-band observations using the
   high-resolution grism. Straight lines correspond to single temperature fits
   to the population distribution, whereas dashed lines plotted in the diagrams
   of knots B1, D1, and c are fits to discrepant measurements (see text).
   }
              \label{excitation}
    \end{figure*}

The excitation diagrams for the various knots are
shown in Fig.~\ref{excitation}. 
Open triangles and open squares correspond to data inferred from the blue and 
red grisms GB and GR, respectively, while filled triangles and squares 
represent data inferred from the H- and K-band observations carried out
at the higher spectral resolution, respectively. 
Straight lines fitted to the various population
distributions show single temperature fits, with values of 
1650K for knot c, 
1815K for knots b and B2,
1855K for knot B4,
1920K for knots B1, B3, D1, D3 and H1,
2030K for knot d and bowNE, 
and 2160K for knot I2. 
Uncertainties in the derived excitation temperatures are about 200~K. 
The temperature range among the various knots is extremely narrow, 
and except for knot c, all data are consistent with a constant excitation 
temperature of $2000 \pm 200$K. 

In a few cases, the population densities inferred from the
various grism settings for a given knot
are not consistent with each other. This is seen in the excitation 
diagrams of Fig.~\ref{excitation}, where most pronounced discrepancies occur 
for knots B1, D1, and c. 
For knot B1, population densities derived from the GB observations are lower
by factors of a few compared to population densities derived from GR and the
H and Ks grisms. For knots D1 and c, population densities derived from the observations
with the GR grism are too low, if compared with the population densities derived from
the GB, H, and K$_\mathrm s$ observations. The discrepancies do not arise from
a larger than assumed reddening towards these knots, which would explain 
the differences in knot B1, but not those in D1 and c. 
They result from relatively severe slit losses during 
phases of poor seeing
{
and from positional differences in the various grism settings, most
pronounced for the GB and GR observations that were taken several
years apart.
However, the temperatures derived from the discrepant population densities 
agree with temperatures derived from the remaining data, for a given knot. 
This is illustrated by the dashed lines shown for knots B1, D1, and c, which are
temperature fits to the discrepant population densities.
Thus, discrepancies do not in any way affect the conclusions 
derived here concerning the H$_2$ excitation, but they severely
hamper a more detailed analysis of the [FeII] emission as discussed below.
}

    \begin{figure}
   \centering
   \includegraphics[width=9cm]{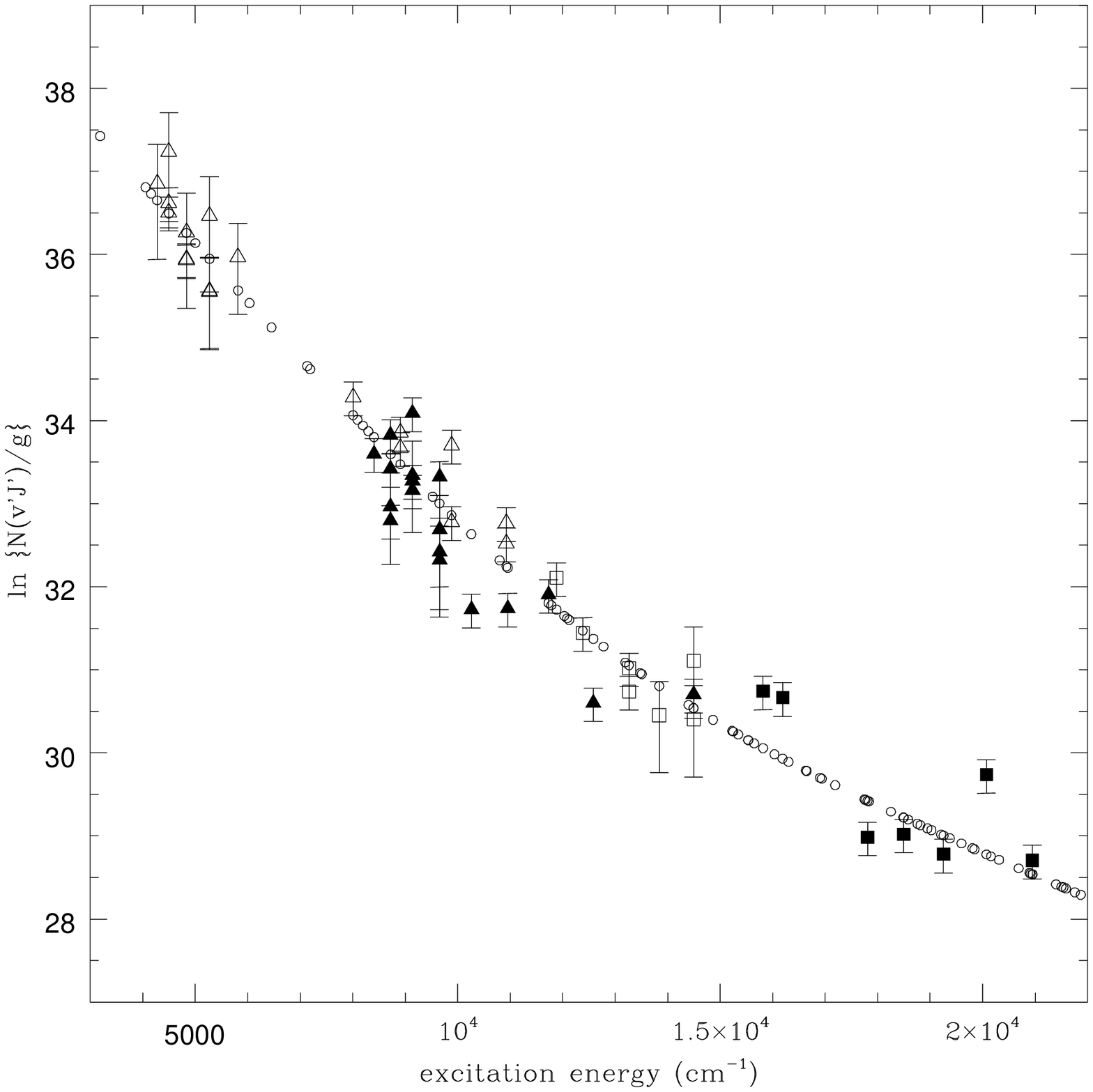}
   \caption{Excitation diagram constructed from the H$_2$ sum spectra.
   Triangles, squares, and hexagrams, respectively, represent population
   distribution among rotational levels of v'=1, 2, 3, and 4.
   Dots represent a theoretical population distribution
   constructed for a thermalized gas at a temperature of 2000K, with a
   second component added at a temperature of 5500K and a column density 
   300 times less than the first.
   }
              \label{sum_exc}
    \end{figure}

%\clearpage

\subsection{Global properties of the H$_2$ emission in HH135/HH136}
\label{sumspectra}
As demonstrated in Fig.~\ref{excitation},
the ro-vibrational population distribution of H$_2$ 
among the various knots is
remarkably similar. It thus makes sense to sum up the individual spectra to increase
the signal-to-noise ratio, to search for fainter H$_2$ emission lines
that are not apparent in the individual spectra. This method simulates spectroscopic
observations of HH135/HH136 with a very large aperture, and is thus
physically meaningful.  The sum spectra, shown in 
Figs.~\ref{spectrum_grf1}--\ref{spectrum_Ks}, show 
a wealth of additional, faint emission lines, particularly in the J-band. 
Some H$_2$ emission lines are identified explicitly in the figures. 
The importance of J-band observations for a proper
characterization of the H$_2$ excitation has been emphasized by Giannini et al.
(\cite{giannini02}), and full use of the additional J-band emission lines is made here.

The H$_2$ line fluxes derived from the sum spectra and the inferred column densities N(v'J') 
are given in Tables~\ref{table_GB}--\ref{table_HKs}. 
The conversion of molecular hydrogen fluxes to population densities was
achieved via the relation $I(\mathrm{v'J'v''J''}) = {hc\over4\pi} \bar{\nu} A(\mathrm{v'J'v''J''})  
N\mathrm{(v'J')}$. 
Intensities $I(\mathrm{v'J'v''J''})$ are inferred from the de-reddened fluxes 
$F(\mathrm{v'J'v''J''})$, using
the solid angles $\Omega$ given in Sect.~\ref{observations}.
The corresponding H$_2$ excitation diagram 
is shown in Fig.~\ref{sum_exc}. The population distribution indicates 
the presence of a thermalized
gas at $T_\mathrm{ex} = 2000$~K, which contains a minor fraction of 0.3\% of hot gas
at $T_\mathrm{ex} = 5500$~K. Dots in the excitation diagram show the modeled population
densities in all levels at such a gas mixture. 

Because of the wealth of [FeII] and other atomic emission lines in the spectra, 
and because of the relatively low spectral resolution, 
the confusion limit is reached in various spectral regions and care needs to be exercised 
when inferring line fluxes of individual emission lines. 
It thus makes sense to calculate a theoretical H$_2$ spectrum 
and compare it with the observed spectrum. This allows  
us to judge whether the fluxes of the molecular hydrogen lines listed 
in Tables~\ref{table_GB}--\ref{table_HKs} are
severely affected from line blends. 
A theoretical H$_2$ spectrum
was constructed for thermalized H$_2$ in a gas mixture of 2000~K and 5500~K with a column 
density ratio of 300:1, using the code
of Gredel \& Dalgarno (\cite{dalgarno}). The theoretical spectra, calculated for a
visual extinction of $A_\mathrm{V} = 2.7$ mag, were normalized to the strength of the
(1,0) S(1) line and convolved with a Voigt profile at the various spectral resolutions
of the observed spectra. The theoretical spectra are reproduced in Figs.~\ref{spectrum_grf1}--
\ref{spectrum_Ks} as bold lines. 

The comparison of the observations with the model spectra show that for single lines
that appear unaffected by blends with atomic lines, the fit is excellent. The H$_2$ line fluxes
in the H- and K-band atmospheric windows are generally very  
well reproduced (cf. Figs.~\ref{spectrum_H} and \ref{spectrum_Ks}). Note that the (1,0) S(1) line
near 2.121 $\mu$m, as well as the $\mathrm{a^4D_{7/2}-a^4F_{9/2}}$ [FeII] line near 1.644
$\mu$m, show broad emission wings. In the heavily crowded J-band spectra, individual H$_2$
lines that are free of blends are very well reproduced as well. Examples are
the (2,0) S(1) and (3,1) S(4) pair near 1.164 $\mu$m (Fig.~\ref{spectrum_grf1})
and the (4,2) S(5), (4,2) S(6), and (2,0) Q(1)-Q(4) lines at 1.21--1.25 $\mu$m 
(Fig.~\ref{spectrum_grf2}). The very good fit of emission lines from v'=2 to v'=4
shows that the two-temperature fit to the H$_2$ population distribution
(Fig.~\ref{sum_exc}) is accurate, despite the flux uncertainties in many 
of the H$_2$ line fluxes that result from line blends. 

Re-forming molecular hydrogen in the cooling, post-shock regions causes
 a pronounced
variation in the population densities of ro-vibrational levels with similar excitation
temperatures,
but different vibrational quantum numbers (e.g., Flower et al. \cite{flower},
their Fig.~6). The effect leads to a pronounced overpopulation of rotational
levels in v'=0, relative to levels in v'=1 of similar excitation energy. 
Emission lines that arise from v'=0 are not available in the present study. 
The rotational population densities in v'=0 are very sensitive to C-type shocks, which 
produce very large column densities in these levels. 
{
C-type shocks have been used to explain the observed H$_2$ in various
outflow sources, such as HH~99 and VLA~1623A (Davis et al. \cite{davis99}) or
in IRAS18151-1208 (Davis et al. \cite{davis04}). Others have used C-type shocks
with magnetic precursors (McCoey et al. \cite{mccoey}; Flower et al. \cite{flower}).
The lack of kinematical data and the lack of observations of H$_2$ emission
from $v=0$ 
{
makes it difficult to judge whether C-type shocks are relevant
for the H$_2$ excitation in HH135/HH136, 
although they must not be ruled out.
}

For a gas mixture of warm molecular gas at 2000K plus a fraction of 0.3\% at 5500~K,
the model calculation produces a total H$_2$ flux of 
$F_\mathrm{tot}$(H$_2) = \Sigma_\mathrm{v'J'v''J''} 
F\mathrm{(v'J'v''J'')} = 14.2 \times F\mathrm{(1301)}$, 
where $F\mathrm{(1301)}$ 
is the flux in the (1,0) S(1)
line and where the summation is carried out over all possible emission lines 
that arise from the E2 cascade in the X$^1\Sigma_\mathrm{g}^+$ electronic
ground state of H$_2$.
The various slit positions employed here sample a total H$_2$
flux of $F_\mathrm{tot}$ (H$_2$) = $1.2 10^{-15}$ W m$^{-2}$, which corresponds
to a total measured luminosity of L(H$_2$) = 0.8 $L_\odot$. From a comparison of
the H$_2$ emission regions sampled by the various slit positions with the overall
morphology of the H$_2$ emission in HH135/HH136, it is estimated that
the total H$_2$ luminosity in HH135/HH136 reaches at least 2 $L_\odot$, but possibly
exceeds this number by factors of a few. 
The total H$_2$ population density, for the two-component mixture assumed
above, is 
$N_\mathrm{tot}(\mathrm{H}_2) = \Sigma_\mathrm{v'J'} N\mathrm{(v'J')} 
= 77 N\mathrm{(1,3)}$, or 
$N_\mathrm{tot}(\mathrm{H}_2) = 5.2 10^{18}$ cm$^{-2}$.  
The scaling factors (14.2 and 77 in the present case) are dependent
on the actual ro-vibrational population distribution 
and show a pronounced dependence on the H$_2$ excitation temperature
(e.g., Gredel \cite{gredel94}).  

Low-mass stars that power molecular outflows such as HH111 and HH212 
have bolometric luminosities of a few tens of solar luminosities 
(Froebrich \cite{froebrich05}),
whereas the H$_2$ luminosities associated with such flows are of the order of
0.1 $L_\odot$. The accretion rates of Class~0 and Class~I objects
are, purportedly,  proportional to the flow luminosities
(Stanke \cite{stanke}; Froebrich et al. \cite{froebrich}), and the
H$_2$ luminosities L(H$_2$) 
seem to increase with increasing source luminosities L(bol).  
The very large luminosity of IRAS11101-5928 suggests the presence 
of a deeply embedded, intermediate- or high-mass star in formation. 
Its luminosity and the luminosity of its H$_2$ outflow define a data point
that is consistent with the regression of L(H$_2$) with L(bol)
from the low-mass protostars. 
This result may be interpreted in terms of
a significantly increased accretion rate in IRAS11101-5829, 
that is, that the intermediate- to high-mass star formation 
witnessed by HH135/HH136 proceeds via accretion as well, but with
significantly increased accretion 
rates, relative to the accretion rates of low-mass protostars.
{
This interpretation is in agreement with Brooks et al. \cite{brooks03},
who suggested that accretion in the luminous IRAS source IRAS 16547-4247
drives the observed, well-collimated outflow
that extends over 1.5 pc. Similarly, Davis et al. \cite{davis04} 
observed a collimated, parsec-scale molecular outflow from IRAS 18151-1208 and
concluded that massive protostars drive collimated jets while in their
earliest stages of evolution. More studies of intermediate- to high-mass
protostars and their outflows are needed to understand if the outflow
properties are scaled-up versions of the low-mass counterparts and 
if high-mass star formation proceeds, in general, via the same
physical processes that govern low-mass star formation.
}

   \begin{figure*}
\sidecaption
   \includegraphics[angle=-90,width=15cm]{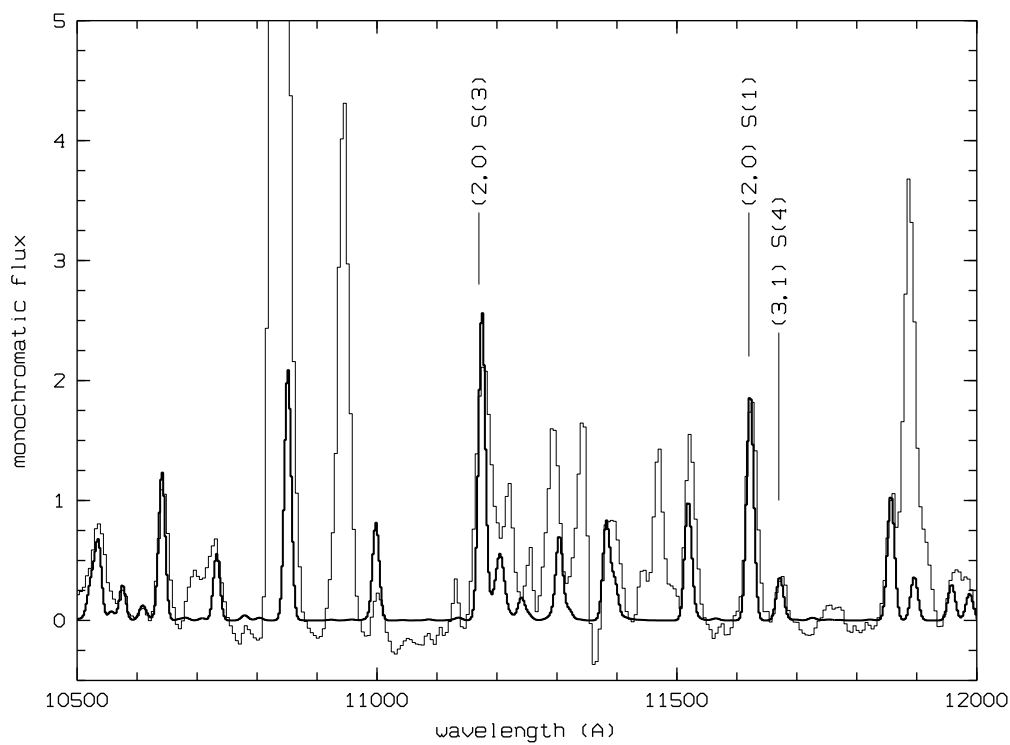}
   \caption{Monochromatic fluxes versus wavelength of the sum spectrum 
    including the model H$_2$ spectrum (bold line). The sum spectrum 
    is obtained from individual spectra using grism GB. A few
H$_2$ lines are identified.
   }
              \label{spectrum_grf1}
    \end{figure*}

   \begin{figure*}
\sidecaption
   \includegraphics[angle=-90,width=15cm]{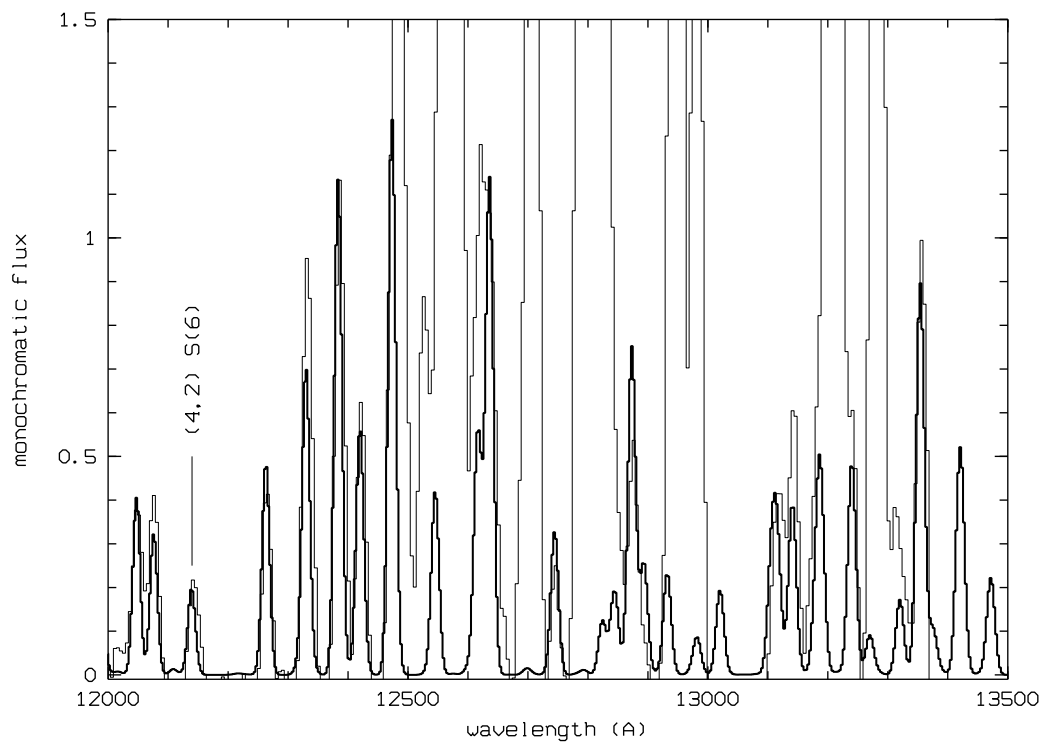}
   \caption{Monochromatic fluxes versus wavelength of the sum spectrum 
    including the model H$_2$ spectrum (bold line). The sum spectrum
    is obtained from individual spectra using grism GB.
The (4,2) S(6) transition of H$_2$ is identified.
}
              \label{spectrum_grf2}
    \end{figure*}

   \begin{figure*}
\sidecaption
   \includegraphics[angle=-90,width=15cm]{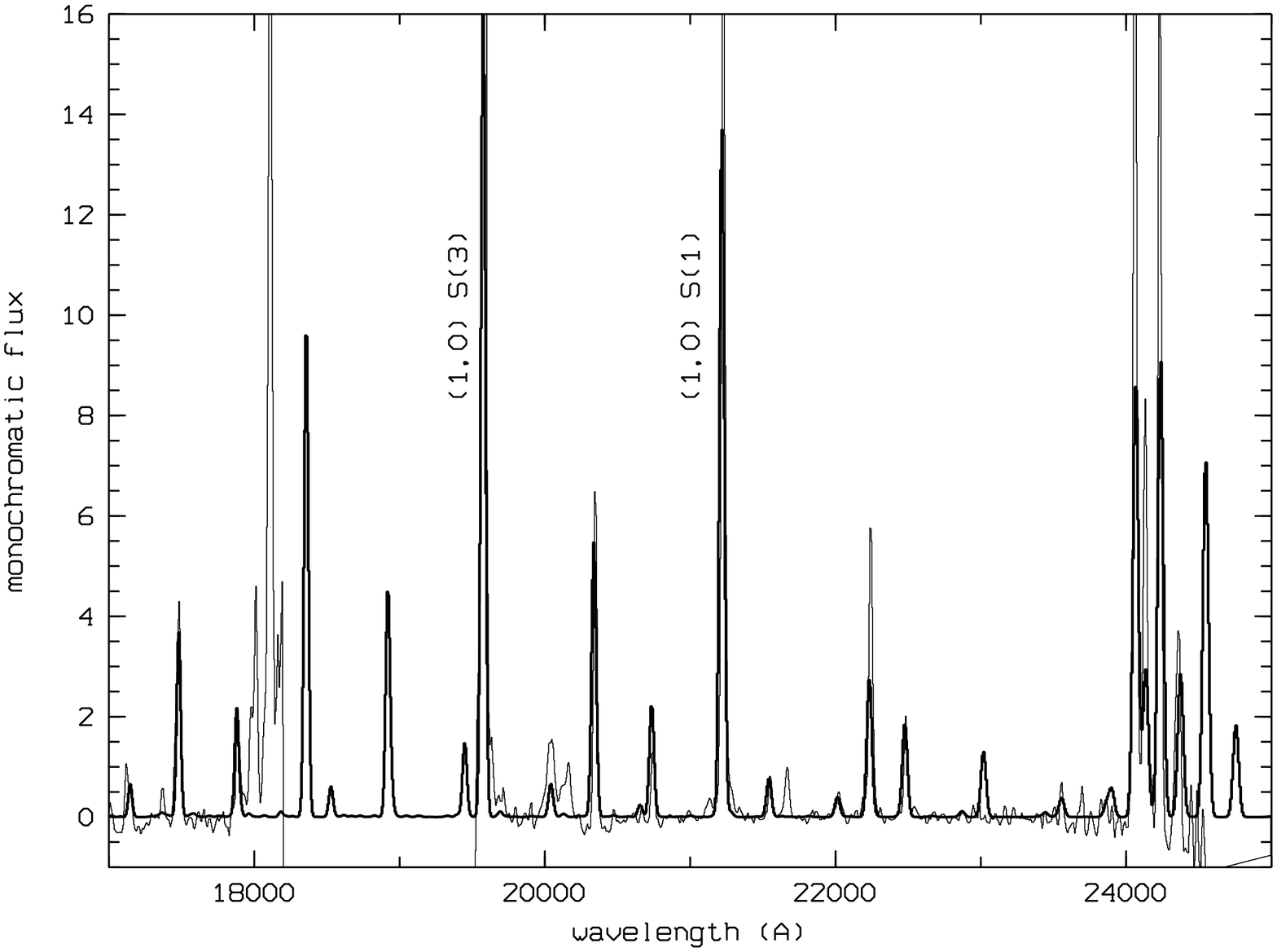}
   \caption{Monochromatic fluxes versus wavelength of the sum spectrum
    including the model H$_2$ spectrum (bold line). The sum spectrum
    is obtained from individual spectra using grism GR.
The (1,0) S(1) and (1,0) S(3) lines of H$_2$ are identified.
   }
              \label{spectrum_grb}
    \end{figure*}

   \begin{figure*}
\sidecaption
   \includegraphics[angle=-90,width=15cm]{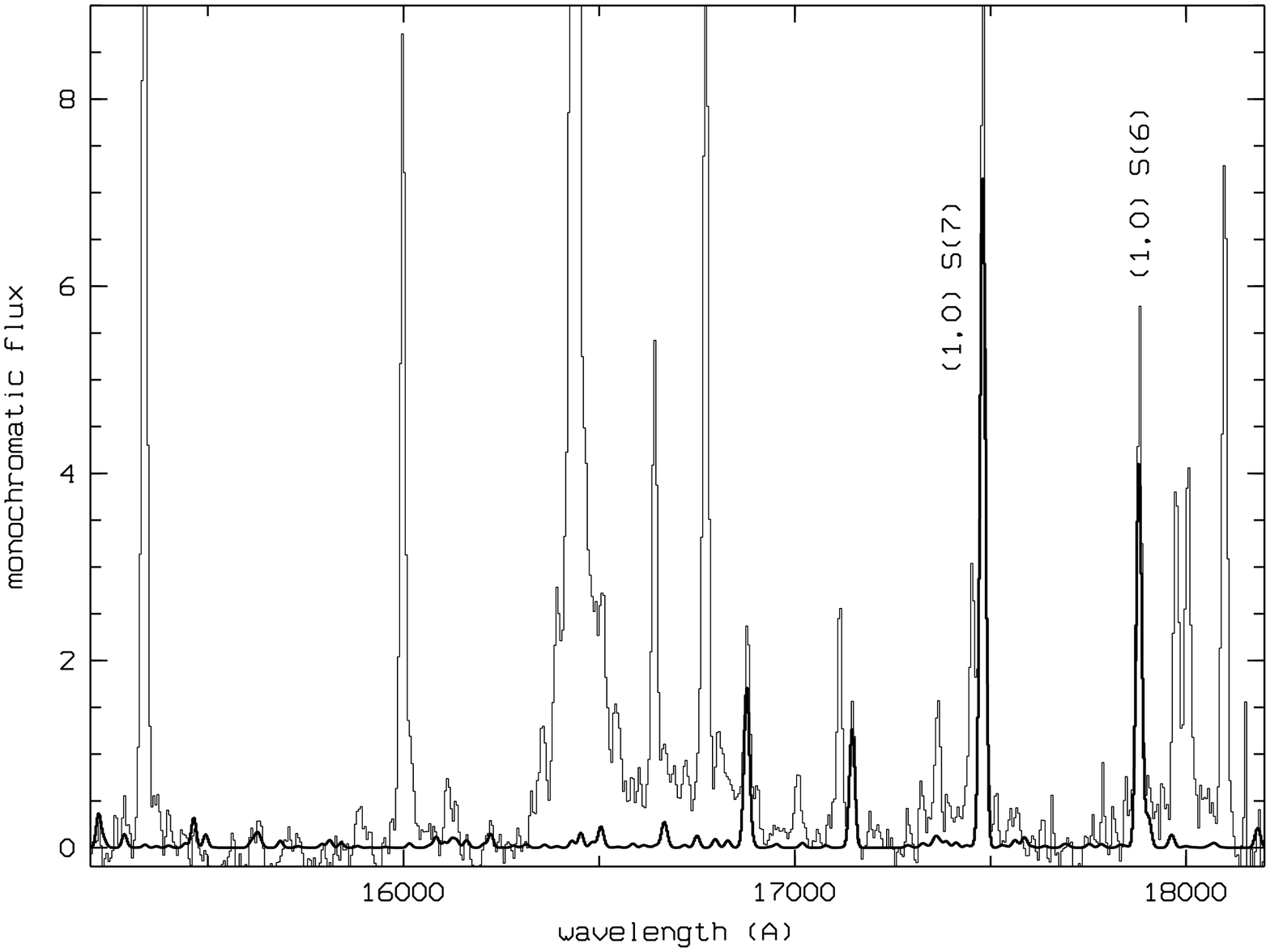}
   \caption{Monochromatic fluxes versus wavelength of the sum spectrum
    including the model H$_2$ spectrum (bold line). The sum spectrum
    is obtained from individual spectra using the HR grism.
The (1,0) S(6) and (1,0) S(7) lines of H$_2$ are identified.
   }
              \label{spectrum_H}
    \end{figure*}

   \begin{figure*}
\sidecaption
   \includegraphics[angle=-90,width=15cm]{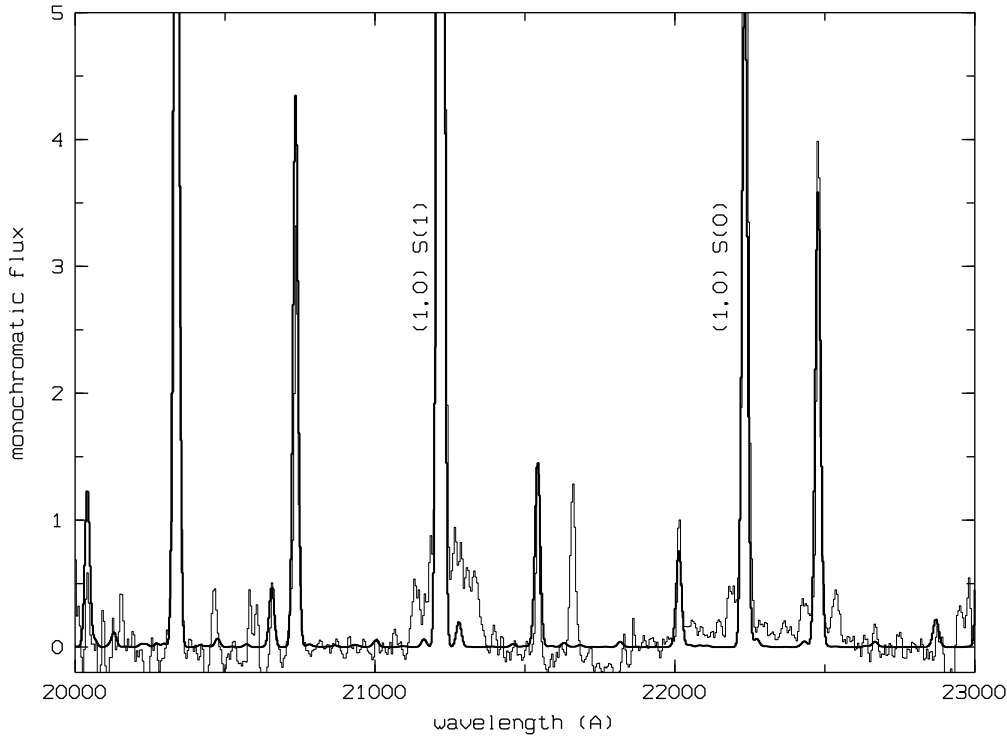}
   \caption{Monochromatic fluxes versus wavelength of the sum spectrum
    including the model H$_2$ spectrum (bold line). The sum spectrum
    is obtained from individual spectra using the HR grism.
The (1,0) S(0) and (1,0) S(1) lines of H$_2$ are identified.
   }
              \label{spectrum_Ks}
    \end{figure*}

\subsection{The atomic emission}
\label{FeII}
The J- and H-band spectra of the various emission line knots
in the HH135/HH136 region are dominated by a wealth of strong emission
lines from [FeII] and from atomic species such as CI, SII, and
NI. Emission from hydrogen recombination lines in the Brackett and
Paschen series are detected as well.
Optical emission from various ions such as [FeII], [FeIII],
[SII], [OI], [OII], [OIII], etc., and from HeI and H$\alpha$ has
been reported by Ogura \& Walsh (\cite{ogura}). Emission from [FeII]
is generally weak in HII regions and planetary nebulae, but rather strong
in supernova remnants and in fast shocks.  

A comprehensive analysis of the [FeII] emission in Herbig-Haro
flows has been presented by Nisini et al. (\cite{nisini}) and by Hartigan, 
Raymond, \& Pierson (\cite{raymond}).  [FeII] emission lines that arise
from a common upper level, such as the 
$\mathrm{a^6D_{9/2}-a^4D_{7/2}} (\lambda 1.257\mu$ m), 
$\mathrm{a^6D_{7/2}-a^4D_{7/2}} (\lambda 1.321\mu$ m),
$\mathrm{a^4F_{9/2}-a^4D_{7/2}} (\lambda 1.644\mu$ m), and
$\mathrm{a^4F_{7/2}-a^4D_{7/2}} (\lambda 1.809\mu$ m) lines, can be used to determine
the reddening towards the line emitting region. 
As pointed out by Hartigan et al. 
(\cite{raymond}), the Einstein A-values of Quinet et al. (\cite{quinet}) 
are probably more accurate than those 
of Nussbaumer \& Storey (\cite{nussbaumer}) because the former result
in a better agreement of the extinction values determined from [FeII] lines, 
compared to the extinction determined from other methods such as the Balmer 
decrement. From the line fluxes of the sum spectra tabulated in 
Tables~\ref{table_atoms1} and \ref{table_atoms2}, we infer a visual
extinction of $A_\mathrm{V}$ = 2.5 - 2.7 mag using Quinet's oscillator strengths. 
This is in excellent agreement with the
values of 2.8--3.2 mag obtained by Ogura \& Walsh (\cite{ogura}) from
their optical observations. 
Nussbaumer's A-values result in visual extinctions of 5.4--5.7 mag, which
are too large if compared with the extinction inferred from the optical data.  
We note, however, that the good agreement between the extinction
derived here and the value of Ogura \& Walsh \cite{ogura}
may be fortuitous, as the extinction
has been derived from [FeII] fluxes in the J- and H-band spectra, which
were taken several years apart, and where positional differences of 
the various slit positions introduce large uncertainties
in the [FeII] line fluxes towards some of the knots in HH135/HH136.

Following Nisini et al. (\cite{nisini}), it is in principle
possible to estimate the electron temperature $T_e$ via the
[FeII]1.257/Pa$\beta$
flux ratio. Our values would 
indicate  an electron temperature of about $T_e \approx 3000$~K.
This value is, however, highly uncertain and probably wrong altogether,
as pointed out by the referee.  The FeII/Pa$\beta$ ratio depends
on both the abundance of iron in the gas phase and on the ionisation 
fraction, two parameters that are not well constrained. In addition,
at such low electron temperatures, the [FeII] levels of the observed
lines, whose excitation energies are above 10\,000~K, are not
populated enough to give rise to the observed, strong emission. 
The low temperature is also not consistent with the fast J-shocks
needed to account for the iron presence in the gas form.
The [FeII]1.257/[CI]0.98 flux ratio, compared with the [FeII]1.247/Pa$\beta$
flux ratio, suggests the presence of a very fast J-type shock, with shock velocities
around $v_s \approx 100$ km s$^{-1}$, assuming a standard gas-phase Fe abundance
of Fe/H = $10^{-6}$. The [FeII]1.644/1.600 ratios are similar to line ratios
observed in other HH objects with [FeII] emission. 
Ratios of 10--15  prevail towards most of the knots in HH135 and HH136
(cf. Tables~\ref{all_H} and \ref{all_GR}) and indicate electron densities around
$n_e = 4000$ cm$^{-3}$ or somewhat below 
(cf. Fig.~7 of Nisini et al. (\cite{nisini})). 
Larger ratios of 20--30 are inferred for knots c and D3, 
and corresponding electron
densities are somewhat below $n_e = 3000$ cm$^{-3}$. 
Flux ratios of [FeII]1.644/1.533 and [FeII]1.644/1.677 of about 10 for all the 
knots indicate electron densities of the order of $n_e = 3500$ cm$^{-3}$. 
The manifold atomic emission lines towards HH135/HH136, together with
the low spectral resolution available here, causes a large number of
line blends (cf. Figs.~\ref{spectrum_grf1} - \ref{spectrum_Ks}), which
prohibits a more comprehensive analysis of the atomic emission.

\section{Conclusions}
\label{conclusions}
The observations presented above and the main conclusions 
are summarized as follows:

\begin{enumerate}
\item 
The images obtained in the (1,0) S(1) line of molecular hydrogen 
reveal the presence of a well-collimated molecular outflow that
extents over a scale of about 1 pc. 
\item
{A number of the H$_2$ emission line knots are associated with
faint, underlying continuum emission.
}
\item
The ro-vibrational excitation temperatures of H$_2$ of the various knots
in the HH135/HH136 flow are remarkably constant,
and are well characterized by a narrow range of 
$T_\mathrm{ex} = 2000 \pm 200$K. 
\item
The molecular part of the shocked gas contains a small fraction of some 
0.3\% of hot H$_2$ at a ro-vibrational excitation temperature of 
$T_\mathrm{ex} = 5500 \pm 200$K. 
\item 
Very strong emission lines from [FeII] occur towards various
knots in HH135/HH136, and emission from HeI, hydrogen recombination lines, 
[CI], [SII], and [NII] are present as well. The [FeII] line ratios indicate 
the presence of a fast J-type shock at a speed of $v_s \approx 100$ km s$^{-1}$.
Electron densities are of the order of $n_e = 3500 - 4000$ cm$^{-3}$ and
electron temperatures are $T_e \approx 3000$~K.  A more comprehensive
analysis of the atomic emission is hampered by the low spectral 
resolution and the occurrence of a very large number of line blends. 
\item 
The ro-vibrational population distribution of H$_2$, together with
the presence of strong [FeII] emission, which is spatially displaced from the
H$_2$ emission, indicate that the emission lines arise form in the cooling 
regions of fast, dissociative J-type shocks, where the [FeII] emission
traces the fast, dissociative parts of the shocks and where the H$_2$ emission
emerges from regions of oblique shocks where the shock speeds are lower. 
The presence of emission from re-forming H$_2$ molecules in the cooling 
post-shock region is not ruled out.
\item
The large H$_2$ luminosity of 2 $L_\odot$ 
suggests that the intermediate- to high-mass protostar that powers the
HH135/HH136 outflow forms via a significantly increased accretion rate, 
compared to the accretion rates of low-mass protostars.

\end{enumerate}

\begin{acknowledgements}
{
It is a great pleasure to thank Drs. Ogura and Sugitani for enlightening
discussions about HH135/HH136 and an invitation to Nagoya City University. 
The useful 
comments of the anonymous referee regarding a better presentation of the
results and an error in the [FeII] analysis are very much appreciated.
}
\end{acknowledgements}

\Online
\appendix

   \begin{figure}
   \centering
\includegraphics[angle=-90,width=12cm]{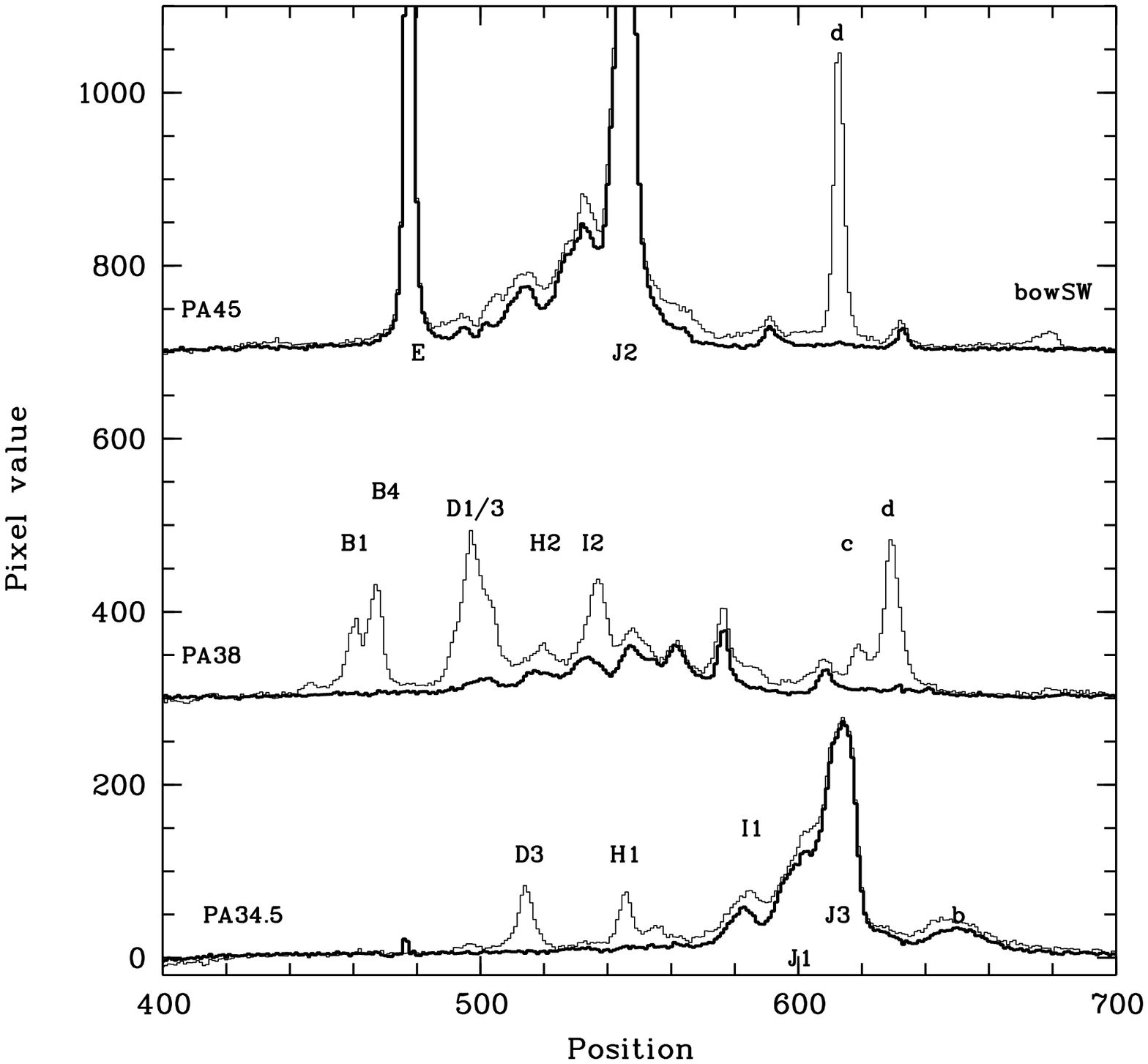}
      \caption{Spatial variation of the (1,0) S(1) H$_2$ emission line flux along the
slit for the three positions PA=34.$^\circ$5, 38$^\circ$, and 45$^\circ$ as inferred 
from the March 1999 observations. Bold lines show 
the variation of the continuum emission. 
              }
         \label{trace_grf}
   \end{figure}

   \begin{figure}
   \centering
\includegraphics[angle=-90,width=12cm]{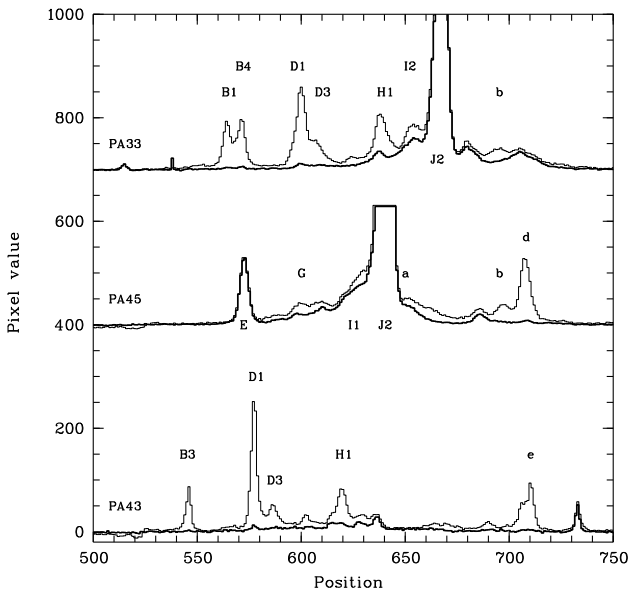}
      \caption{Spatial variation of the (1,0) S(1) H$_2$ emission line flux along the
slit at the three positions PA=33$^\circ$, 45$^\circ$, and 43$^\circ$ as inferred from
the February 2004 observations.  Bold lines show the variation of 
the continuum emission. 
              }
         \label{trace_ks}
   \end{figure}

   \begin{figure}
   \centering
\includegraphics[angle=-90,width=10cm]{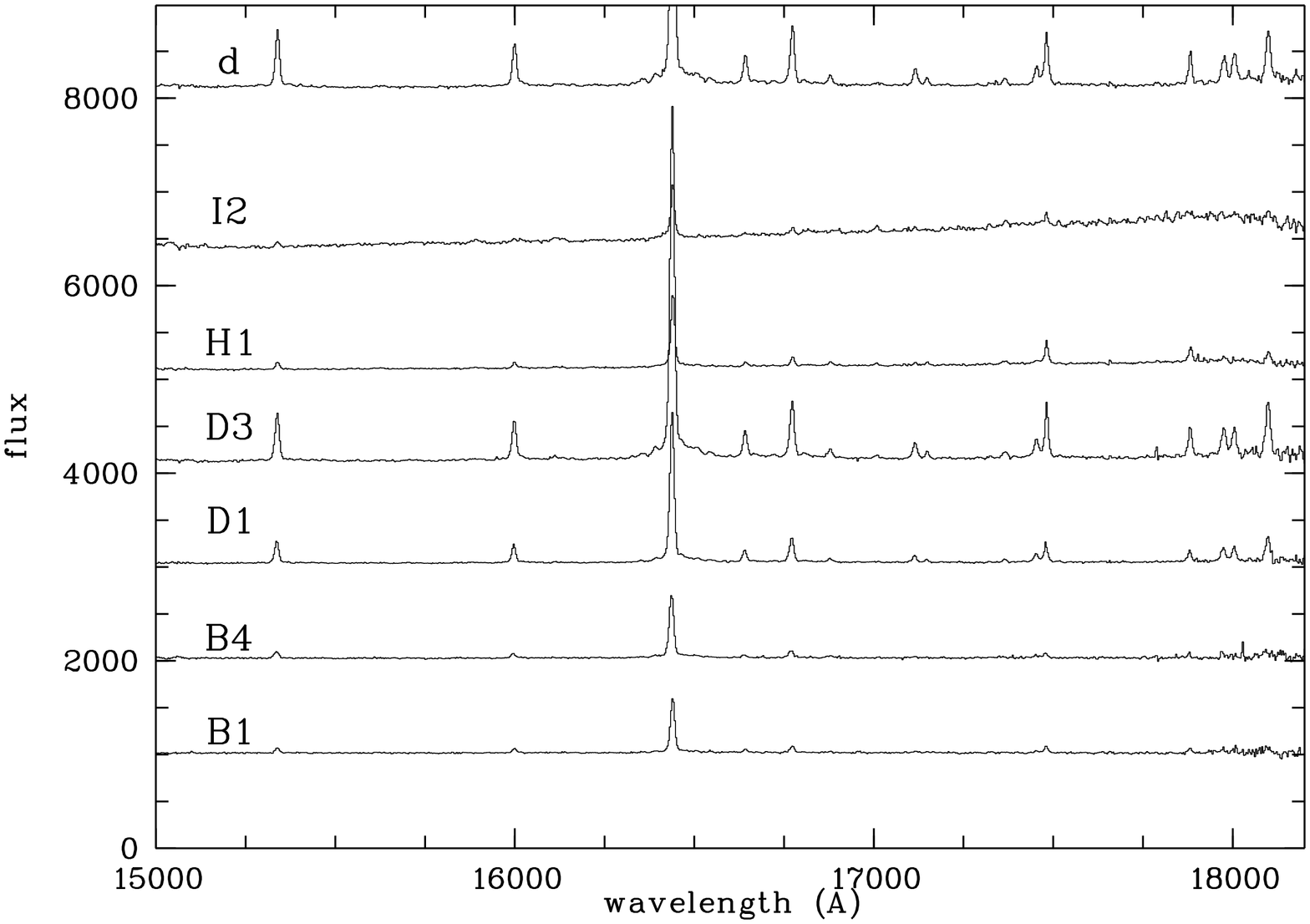}
      \caption{
	      H-band spectra of individual knots in the HH135 flow. Spectra are shifted
	      by units of 1000, 2000, 3000, 4000, 5000, 6000, and 8000 for knots
	      HH135-B1, B4, D1, D3, H1, I2, and HH136-d, 
	      respectively, along the ordinate.
              }
         \label{spectra_H}
   \end{figure}

   \begin{figure}
   \centering
\includegraphics[angle=-90,width=10cm]{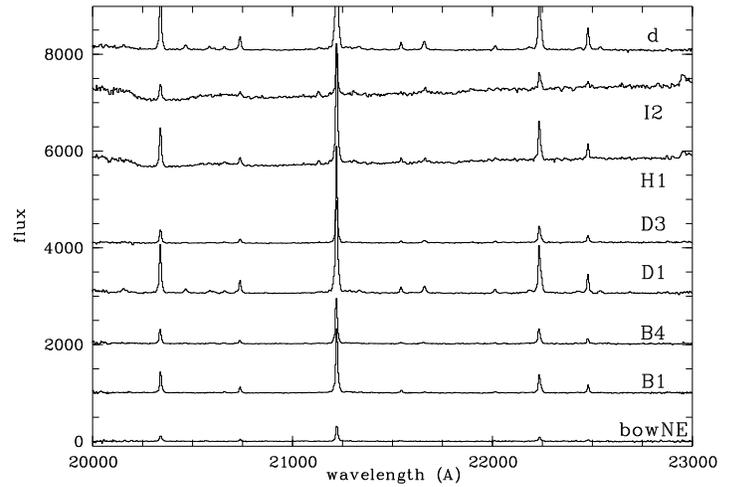}
      \caption{
	      K-band spectra of individual knots in the HH135 flow. The ordinate is valid
	      for bowNE only. Spectra are shifted by units of 1000, 2000, 3000, 4000, 
	      5000, 6000, and 8000 for knots HH135-B1, B4, D1, D3, H1, I2, and HH136-d, 
	      respectively, along the ordinate.
              }
         \label{spectra_ks}
   \end{figure}

\clearpage

   \begin{table*}
      \caption[]{
Individual measurements of emission lines towards the various knots in
HH135 and HH136, using the low resolution spectra obtained with grism GB.
      }
         \label{all_GB}
\centering
     $$ 
         \begin{array}{p{0.25\linewidth}rrrrrrrrrrrrrrrrrrrr}
            \hline \noalign{\smallskip} \hline
            Wavelength ($\mu$m)    
&  \multicolumn{10}{c}{\mathrm{Line\ flux}(10^{-20} \mathrm{W\ m}^{-2})} \\
            \noalign{\smallskip} \hline \noalign{\smallskip}
& \multicolumn{3}{c}{PA33} & \multicolumn{5}{c}{PA43} & \multicolumn{2}{c}{PA65}\\
& \multicolumn{3}{c}{\hrulefill}& \multicolumn{5}{c}{\hrulefill} & \multicolumn{2}{c}{\hrulefill}\\
& B1 & B4 & D1 & B4 & D1 & H1 & e & a & B1 & B2\\
            \noalign{\smallskip} \hline \noalign{\smallskip}

0.953 &   0.4 &   1.4 &   1.0 &   0.3 &   0.6 &   - &   - &   3.5 &   0.6 &   0.6 \\
0.983 &   0.2 &   0.3 &   0.3 &   0.1 &   0.3 &   0.1 &   0.3 &   0.7 &   0.3 &   0.3 \\
0.986 &   0.5 &   0.9 &   0.8 &   0.4 &   0.6 &   0.2 &   0.8 &   1.9 &   0.7 &   0.6 \\
1.006 &   0.2 &   0.5 &   0.6 &   0.2 &   0.4 &   0.2 &   0.1 &   1.3 &   0.4 &   0.3 \\
1.0293 &   0.5 &   1.7 &   1.8 &   0.4 &   1.8 &   0.1 &   0.3 &   4.6 &   0.7 &   0.6 \\
1.033 &   1.3 &   3.8 &   4.2 &   1.1 &   2.6 &   0.4 &   0.8 &  10.9 &   1.6 &   1.5 \\
1.038 &   0.3 &   0.8 &   0.8 &   0.2 &   0.8 &   0.1 &   0.1 &   2.0 &   0.3 &   0.4 \\
1.041 &   0.5 &   1.2 &   2.1 &   0.2 &   1.2 &   0.2 &   0.2 &   4.9 &   0.4 &   0.4 \\
1.053 &   0.1 &   - &   0.4 &   0.1 &   - &   - &   0.1 &   0.3 &   - &   - \\
1.064 &   0.2 &   0.3 &   0.2 &   - &   0.1 &   - &   0.2 &   0.5 &   - &   0.2 \\
1.072 &   0.2 &   0.4 &   0.3 &   - &   - &   - &   0.2 &   0.9 &   0.3 &   0.2 \\
1.084 &   3.3 &  10.2 &   8.7 &   3.4 &   6.8 &   1.1 &   1.8 &  22.7 &   4.4 &   3.9 \\
1.095 &   0.3 &   1.0 &   1.1 &   0.5 &   1.0 &   0.4 &   0.5 &   2.4 &   0.4 &   0.5 \\
1.163 &   0.2 &   0.4 &   0.4 &   0.1 &   0.3 &   0.1 &   - &   0.6 &   0.3 &   0.3 \\
1.189 &   0.4 &   1.1 &   1.2 &   0.4 &   0.8 &   0.2 &   0.3 &   2.5 &   0.8 &   0.6 \\
1.233 &   0.1 &   0.2 &   0.2 &   0.1 &   0.2 &   0.1 &   0.3 &   0.5 &   0.2 &   0.1 \\
1.239 &   0.1 &   0.3 &   0.3 &   0.2 &   0.2 &   0.1 &   0.4 &   0.4 &   0.3 &   0.3 \\
1.248 &   0.3 &   0.7 &   0.9 &   0.3 &   0.9 &   0.2 &   0.3 &   1.7 &   0.6 &   0.5 \\
1.257 &   8.5 &  20.0 &  22.4 &   7.7 &  17.1 &   3.8 &   5.9 &  39.5 &  11.1 &   9.8 \\
1.271 &   0.4 &   1.2 &   1.5 &   0.4 &   1.1 &   0.1 &   0.1 &   2.6 &   0.5 &   0.4 \\
1.279 &   0.7 &   2.0 &   2.4 &   0.9 &   2.2 &   0.3 &   0.2 &   4.4 &   0.9 &   0.7 \\
1.282 &   0.7 &   2.4 &   3.8 &   0.6 &   2.8 &   1.4 &   1.0 &   6.3 &   1.2 &   1.2 \\
1.295 &   1.1 &   2.9 &   3.7 &   1.0 &   2.8 &   0.5 &   0.5 &   6.6 &   1.4 &   1.1 \\
1.298 &   0.3 &   0.8 &   0.9 &   0.2 &   0.7 &   0.1 &   - &   1.7 &   0.3 &   0.3 \\
1.321 &   2.5 &   5.7 &   7.3 &   2.2 &   5.6 &   1.3 &   1.8 &  12.0 &   3.4 &   2.9 \\
1.328 &   0.7 &   1.8 &   2.1 &   0.5 &   1.6 &   0.2 &   0.3 &   3.7 &   0.8 &   0.5 \\
1.534 &   1.7 &   4.2 &   6.6 &   1.4 &   4.9 &   1.1 &   0.8 &   8.4 &   2.0 &   1.6 \\
1.600 &   1.2 &   3.1 &   5.5 &   1.0 &   4.0 &   0.8 &   0.6 &   - &   1.5 &   1.3 \\
            \noalign{\smallskip}
            \hline
         \end{array}
     $$ 
\begin{list}{}{}
\item[$^{\mathrm{a}}$] Flux uncertainties are 10\% in general, but increase to
about 50\% for $1.10 \le \lambda \le 1.18 \mu$m.
\end{list}
   \end{table*}

   \begin{table*}
      \caption[]{
Individual K-band measurements of emission lines towards the various knots in
HH135 and HH136, using the medium resolution spectra obtained with grism HR. 
      }
         \label{all_ks}
\centering
     $$ 
         \begin{array}{p{0.25\linewidth}rrrrrrrrrrrrrrrrrrrrr}
            \hline \noalign{\smallskip} \hline
            Wavelength ($\mu$m)    
&  \multicolumn{15}{c}{\mathrm{Line\ flux}(10^{-20} \mathrm{W\ m}^{-2})} \\
            \noalign{\smallskip} \hline \noalign{\smallskip}
& \multicolumn{3}{c}{PA65} & \multicolumn{2}{c}{PA45} & \multicolumn{3}{c}{PA43}
& \multicolumn{7}{c}{PA33} \\
& \multicolumn{3}{c}{\hrulefill} & \multicolumn{2}{c}{\hrulefill} & \multicolumn{3}{c}{\hrulefill}
& \multicolumn{7}{c}{\hrulefill} \\
& B2 & B1 & bowNE & d & c & D1 & D3 & B3 & bowNE & b & H1 & I2 & D1 & B4 & B1 \\
            \noalign{\smallskip} \hline \noalign{\smallskip}

2.034 &  14.2 &   8.4 &   2.9 &  17.0 &   8.4 &  21.3 &   6.6 &   6.5 &   2.9 &   7.6 &  17.9 &   - &  37.1 &  12.5 &  10.7 \\
2.046 &   0.9 &   - &   - &   1.0 &   0.2 &   3.6 &   - &   0.3 &   - &   - &   - &   - &   7.8 &   1.4 &   - \\
2.074 &   2.6 &   2.1 &   0.6 &   2.9 &   0.8 &   5.2 &   1.4 &   1.1 &   0.6 &   1.6 &   2.9 &   - &   7.3 &   2.9 &   2.3 \\
2.122 &  39.5 &  26.3 &   6.8 &  47.9 &  21.0 &  59.2 &  19.0 &  19.5 &   7.3 &  22.7 &  48.4 &  19.6 & 103.3 &  36.3 &  29.9 \\
2.154 &   1.4 &   1.0 &   0.4 &   1.7 &   0.8 &   2.5 &   0.8 &   0.5 &   0.3 &   1.7 &   1.0 &   - &   4.0 &   1.5 &   1.1 \\
2.166 &   0.6 &   0.4 &   - &   0.2 &   0.5 &   3.6 &   1.4 &   0.7 &   - &   1.2 &   - &   1.5 &   7.1 &   1.6 &   - \\
2.201 &   0.7 &   0.5 &   0.2 &   0.5 &   0.4 &   1.2 &   0.3 &   0.3 &   0.4 &   - &   0.4 &   - &   1.5 &   0.9 &   0.6 \\
2.223 &  11.1 &   7.4 &   1.7 &  12.5 &   5.7 &  20.8 &   6.6 &   6.5 &   1.6 &   6.6 &  13.8 &   5.7 &  37.7 &  12.4 &   8.0 \\
2.247 &   3.4 &   2.4 &   0.6 &   4.5 &   3.1 &   6.6 &   2.4 &   1.9 &   1.0 &   2.1 &   4.2 &   2.3 &  10.5 &   3.5 &   2.9 \\
            \noalign{\smallskip}
            \hline
         \end{array}
     $$ 
\begin{list}{}{}
\item[$^{\mathrm{a}}$] Flux uncertainties are 10\% in general, 
but increase to about 50\% for $\lambda \le 2.05 \mu$m.
\end{list}
   \end{table*}

   \begin{table*}
      \caption[]{
Individual H-band measurements of emission lines towards the various knots in
HH135 and HH136, using the medium resolution spectra obtained with grism HR. 
      }
         \label{all_H}
\centering
     $$ 
         \begin{array}{p{0.25\linewidth}rrrrrrrrrrrrrrrrrrrrrr}
            \hline \noalign{\smallskip} \hline
            Wavelength ($\mu$m)    
&  \multicolumn{14}{c}{\mathrm{Line\ flux}(10^{-20} \mathrm{W\ m}^{-2})} \\
            \noalign{\smallskip} \hline \noalign{\smallskip}
& \multicolumn{3}{c}{PA33} & \multicolumn{4}{c}{PA43} & \multicolumn{3}{c}{PA33}
& \multicolumn{2}{c}{PA45} & \multicolumn{2}{c}{PA65} \\
& \multicolumn{3}{c}{\hrulefill} & \multicolumn{4}{c}{\hrulefill} & \multicolumn{3}{c}{\hrulefill}
& \multicolumn{2}{c}{\hrulefill} & \multicolumn{2}{c}{\hrulefill} \\
& B1B4 & D1D3 & H1 & B3 & D1 & H1 & e & B4 & D1 & H1 & c & d & B1 & B2\\
            \noalign{\smallskip} \hline \noalign{\smallskip}

1.534 &  12.5 &  18.7 &   1.9 &   3.6 &   9.8 &   3.4 &   1.9 &   3.0 &   8.3 &   3.1 &   2.8 &  23.5 &   2.0 &   6.2 \\
1.600 &   9.3 &  12.7 &   1.1 &   2.3 &   7.4 &   1.9 &   1.2 &   1.4 &   5.3 &   2.1 &   1.5 &  15.7 &   1.7 &   4.8 \\
1.644 &  99.3 & 159.4 &  24.4 &  32.2 &  71.0 &  30.7 &  20.9 &  29.2 &  64.9 &  29.5 &  26.7 & 155.4 &  23.4 &  65.3 \\
1.664 &   6.4 &   7.8 &   0.9 &   1.9 &   5.3 &   1.4 &   0.7 &   0.6 &   3.1 &   1.3 &   1.3 &  10.9 &   1.2 &   4.3 \\
1.677 &  10.5 &  17.3 &   2.0 &   3.2 &   8.9 &   2.8 &   1.5 &   2.0 &   7.3 &   2.6 &   2.0 &  19.7 &   2.1 &   6.8 \\
1.688 &   2.0 &   1.5 &   1.3 &   - &   2.0 &   1.5 &   1.1 &   0.5 &   0.6 &   1.3 &   1.6 &   3.5 &   0.3 &   1.7 \\
1.712 &   2.8 &   3.5 &   0.7 &   - &   2.2 &   1.0 &   0.1 &   0.2 &   1.6 &   0.6 &   0.6 &   6.1 &   0.4 &   1.3 \\
1.715 &   1.2 &   0.6 &   - &   - &   0.6 &   0.6 &   0.7 &   - &   0.3 &   0.5 &   1.0 &   2.2 &   - &   0.7 \\
1.736 &   - &   - &   - &   - &   - &   - &   - &   - &   - &   - &   - &   2.4 &   - &   - \\
1.745 &   2.3 &   4.5 &   - &   - &   2.5 &   0.6 &   0.3 &   - &   2.4 &   0.8 &   - &   6.9 &   0.6 &   2.1 \\
1.748 &  11.0 &  12.6 &   3.6 &   2.2 &   5.4 &   5.5 &   5.6 &   1.3 &   4.3 &   5.5 &   6.5 &  14.2 &   2.0 &   6.9 \\
1.788 &   5.9 &   6.8 &   - &   - &   2.8 &   3.6 &   3.8 &   - &   2.3 &   6.0 &   4.1 &   7.2 &   1.2 &   4.2 \\
1.798 &   4.8 &   5.5 &   - &   - &   3.8 &   - &   - &   - &   3.2 &   3.3 &   0.5 &   5.7 &   - &   3.4 \\
1.800 &   1.8 &   0.6 &   - &   - &   3.8 &   - &   - &   - &   0.8 &   - &   - &   4.1 &   - &   2.6 \\
1.810 &   5.6 &   9.5 &   - &   - &   5.6 &   - &   - &   - &   4.2 &   - &   - &   8.1 &   - &   4.3 \\
            \noalign{\smallskip}
            \hline
         \end{array}
     $$ 
\begin{list}{}{}
\item[$^{\mathrm{a}}$] Flux uncertainties are 10\% in general. 
\end{list}
   \end{table*}

   \begin{table*}
      \caption[]{
Individual H-band measurements of emission lines towards the various knots in
HH135 and HH136, using the low resolution spectra obtained with grism GR. 
      }
         \label{all_GR}
\centering
     $$ 
         \begin{array}{p{0.25\linewidth}rrrrrrrrrrrrrrrrrrrrrr}
            \hline \noalign{\smallskip} \hline
            Wavelength ($\mu$m)    
&  \multicolumn{15}{c}{\mathrm{Line\ flux}(10^{-20} \mathrm{W\ m}^{-2})} \\
            \noalign{\smallskip} \hline \noalign{\smallskip}
& \multicolumn{6}{c}{PA38}& \multicolumn{1}{c}{PA45} & \multicolumn{2}{c}{PA65}
& \multicolumn{3}{c}{PA34.5} & \multicolumn{3}{c}{PA34.5}\\
& \multicolumn{6}{c}{\hrulefill}& \multicolumn{1}{c}{\hrulefill} & \multicolumn{2}{c}{\hrulefill}
& \multicolumn{3}{c}{\hrulefill} & \multicolumn{3}{c}{\hrulefill}\\
& D1 & B1 & B4 & D1 & d & c & d & bowNE & B2 & D3 & H1 & b & D3 & H1 & b\\
            \noalign{\smallskip} \hline \noalign{\smallskip}

1.534 &   4.3 &   2.8 &   8.0 &  11.6 &  20.2 &   0.5 &  15.0 &   0.2 &   4.5 &   0.9 &   - &   - &   1.2 &   3.0 &   0.1 \\
1.600 &   3.0 &   1.8 &   5.5 &   8.7 &  14.2 &   0.3 &  11.1 &   - &   3.6 &   0.5 &   - &   - &   0.4 &   1.9 &   - \\
1.644 &  27.3 &  18.6 &  54.0 &  91.7 & 112.2 &   8.4 &  82.8 &   2.4 &  36.7 &   9.3 &   6.8 &   3.3 &  12.1 &  25.0 &   2.2 \\
1.664 &   1.6 &   0.8 &   3.5 &   5.0 &   7.4 &   - &   6.1 &   - &   1.7 &   0.3 &   - &   - &   0.4 &   0.9 &   - \\
1.678 &   3.3 &   1.8 &   6.1 &  10.7 &  15.2 &   0.6 &  11.4 &   0.2 &   3.8 &   - &   - &   - &   0.7 &   2.3 &   - \\
1.712 &   0.8 &   0.4 &   1.4 &   2.3 &   4.3 &   0.1 &   3.0 &   0.2 &   1.0 &   0.2 &   - &   - &   0.2 &   0.7 &   - \\
1.748 &   3.7 &   3.3 &   7.9 &  12.1 &  18.3 &   2.0 &  14.9 &   1.2 &   4.7 &   1.9 &   - &   - &   1.6 &   6.4 &   1.2 \\
1.789 &   1.4 &   1.7 &   3.2 &   5.8 &   5.3 &   1.4 &   6.3 &   0.7 &   1.8 &   0.8 &   - &   - &   2.2 &   5.1 &   1.9 \\
1.798 &   3.1 &   1.5 &   4.8 &  13.2 &  13.2 &   1.4 &   9.4 &   0.4 &   3.9 &   2.5 &   - &   - &   5.6 &   5.2 &   - \\
1.810 &  21.0 &  14.9 &  34.8 &  76.9 &  85.7 &   6.0 &  72.3 &   2.2 &  26.7 &  10.8 &   - &   - &  15.5 &  17.5 &   - \\
1.958 &  22.0 &  24.3 &  37.8 &  74.6 &  84.3 &   8.4 &  81.9 &   7.4 &  22.5 &   7.7 &   - &   - &   8.6 &  36.6 &   - \\
2.034 &   4.7 &   5.5 &  10.3 &  19.8 &  18.4 &   2.8 &  17.2 &   1.7 &   7.2 &   2.9 &   5.7 &   2.9 &   6.4 &  14.2 &   5.8 \\
2.047 &   0.7 &   - &   - &   - &   - &   - &   5.6 &   - &   0.4 &   - &   - &   - &   0.7 &   - &   - \\
2.074 &   0.7 &   0.9 &   2.0 &   2.8 &   2.6 &   - &   3.4 &   0.3 &   1.3 &   0.2 &   - &   - &   - &   5.5 &   - \\
2.122 &  12.1 &  14.5 &  29.9 &  53.0 &  48.5 &   7.2 &  47.1 &   3.8 &  19.1 &   7.1 &  10.4 &   7.3 &  12.4 &  27.7 &  12.7 \\
2.155 &   0.5 &   0.6 &   1.2 &   1.9 &   3.2 &   0.7 &   1.0 &   0.3 &   0.6 &   0.2 &   - &   - &   0.5 &   1.7 &   - \\
2.167 &   1.1 &   - &   0.8 &   3.3 &   4.1 &   0.3 &   1.4 &   - &   - &   0.4 &   - &   - &   0.7 &   2.9 &   - \\
2.202 &   0.3 &   0.3 &   0.7 &   1.1 &   1.3 &   0.3 &   0.8 &   0.2 &   - &   - &   - &   - &   - &   - &   - \\
2.224 &   4.3 &   4.5 &   9.7 &  18.4 &  17.6 &   2.1 &  14.9 &   1.2 &   5.8 &   2.4 &   4.0 &   3.2 &   4.9 &   8.5 &   3.7 \\
2.248 &   1.3 &   1.6 &   3.0 &   5.6 &   6.0 &   1.0 &   4.9 &   0.6 &   1.5 &   1.0 &   1.9 &   1.6 &   2.0 &   3.7 &   1.3 \\
2.355 &   1.3 &   0.5 &   0.7 &   1.9 &   2.2 &   0.4 &   1.7 &   0.5 &   0.5 &   0.6 &   - &   - &   - &   - &   - \\
2.406 &  13.5 &  15.4 &  31.1 &  58.9 &  50.9 &   7.7 &  53.3 &   4.1 &  21.6 &   8.9 &  13.1 &   7.9 &  16.4 &  26.7 &  15.3 \\
2.413 &   5.8 &   8.7 &  15.2 &  36.5 &  25.8 &   4.0 &  26.8 &   2.0 &  10.0 &   4.1 &   6.3 &   4.3 &   7.9 &  12.3 &   7.6 \\
2.423 &  14.0 &  15.4 &  31.6 &  59.2 &  50.9 &   7.2 &  53.1 &   4.1 &  21.0 &   7.2 &  12.4 &   9.1 &  13.8 &  29.2 &  16.7 \\
2.437 &   - &   5.0 &   9.0 &  16.9 &  15.3 &   3.3 &  17.6 &   1.5 &   6.2 &   2.4 &   - &   - &   - &   - &   - \\
2.453 &   - &   - &   1.3 &   - &   - &   - &   4.5 &   - &   2.3 &   - &   - &   - &   - &   - &   - \\
2.475 &   - &   2.1 &   3.8 &  12.9 &   4.9 &   - &   4.9 &   - &   - &   - &   - &   - &   - &   - &   - \\
2.498 &   - &   3.7 &   9.7 &  25.4 &  12.5 &   - &  12.5 &   - &   - &   - &   - &   - &   - &   - &   - \\
            \noalign{\smallskip}
            \hline
         \end{array}
     $$ 
\begin{list}{}{}
\item[$^{\mathrm{a}}$] Flux uncertainties are 10\% in general, 
but increase to about 50\% for $1.9 \le \lambda \le 2.05 \mu$m 
and for $\lambda \ge 2.43 \mu$m.
\end{list}
   \end{table*}

   \begin{table}
      \caption[]{
H$_2$ line fluxes and column densities in ro-vibrational levels v'J',
as measured from the low-resolution sum spectrum obtained with the 
blue grism GB.
      }
         \label{table_GB}
\centering
     $$ 
         \begin{array}{p{0.25\linewidth}rrrr}
            \hline
            \noalign{\smallskip}
            Line transition      &  \mathrm{Wavelength} & 
\mathrm{Flux} & N\mathrm{(v'J')}\\
&  \mu \mathrm{m} & {10^{-20} \mathrm{W\ m}^{-2}} & 10^{14} \mathrm{cm}^{-2}\\
            \noalign{\smallskip}
            \hline
            \noalign{\smallskip}

(2,0)S( 9)  &     1.054  &      156 &        7 (        1 )\\ 
(2,0)S( 7)  &     1.064  &      128 &        5 (        1 )\\ 
(2,0)S( 6)  &     1.073  &      141 &        6 (        1 )\\ 
(2,0)S( 4)  &     1.100  &       82 &        4 (        1 )\\ 
(2,0)S( 3)  &     1.117  &      379 &       19 (       10 )\\ 
(2,0)S( 2)  &     1.138  &      225 &       14 (        7 )\\ 
(3,1)S( 5)  &     1.152  &      188 &        4(        2 )\\ 
(2,0)S( 1)  &     1.162  &      288 &       22 (       11 )\\ 
(3,1)S( 4)  &     1.167  &       55 &        1 (        0.5 )\\ 
(3,1)S( 3)$^a$  &     1.186  &      170 &        4 (        1 )\\ 
(4,2)S( 9)  &     1.196  &       62 &        1 (        0.5 )\\ 
(4,2)S( 8)  &     1.199  &       59 &        1 (        0.5 )\\ 
(4,2)S( 7)  &     1.205  &       67 &        1 (        0.5 )\\ 
(4,2)S( 2)  &     1.285  &       56 &        1 (        0.5 )\\ 
(4,2)S( 6)  &     1.214  &       27 &        0.4 (     0.2 )\\ 
(4,2)S( 5)  &     1.226  &       69 &        1 (        0.5 )\\ 
(3,1)S( 1)  &     1.233  &      156 &        5 (        1 )\\ 
(2,0)Q( 2)  &     1.242  &       95 &       11 (        2 )\\ 
(2,0)Q( 3)  &     1.247  &      465 &       57 (       11 )\\ 
(2,0)Q( 4)  &     1.255  &      251 &       32 (        6 )\\ 
(2,0)Q( 5)  &     1.264  &      420 &       55 (       11 )\\ 
(2,0)Q( 7)  &     1.287  &      114 &       16 (        3 )\\ 
(3,1)Q( 1)  &     1.314  &      138 &        5 (        1 )\\ 
(4,2)S( 1)  &     1.312  &      125 &        3 (        1 )\\ 
            \noalign{\smallskip}
            \hline
         \end{array}
     $$ 
\begin{list}{}{}
\item[$^{\mathrm{a}}$] Possible blend with [PI].
\end{list}
   \end{table}

   \begin{table}
      \caption[]{
H$_2$ line fluxes and column densities in ro-vibrational levels v'J',
as measured from the low-resolution sum spectrum obtained with the 
red grism GR.
      }
         \label{table_GR}
\centering
     $$ 
         \begin{array}{p{0.25\linewidth}rrrr}
            \hline
            \noalign{\smallskip}
            Line transition      &  \mathrm{Wavelength} & \mathrm{Flux} 
& N\mathrm{(v'J')} \\
&  \mu\mathrm{m} & {10^{-20} \mathrm{W\ m}^{-2}} & 10^{14} \mathrm{cm}^{-2}\\
            \noalign{\smallskip}
            \hline
            \noalign{\smallskip}
	    
(1,0)S( 9)  &     1.688  &      510 &       65 (       13 )\\ 
(1,0)S( 8)  &     1.715  &      700 &       65 (       13 )\\ 
(1,0)S( 7)  &     1.748  &     2798 &      207 (       41 )\\ 
(1,0)S( 3)  &     1.958  &    17948 &     1056 (      528 )\\ 
(1,0)S( 2)  &     2.034  &     3016 &      195 (       97 )\\ 
(2,1)S( 3)  &     2.074  &      897 &       41 (       20 )\\ 
(1,0)S( 1)  &     2.122  &     8657 &      669 (      134 )\\ 
(2,1)S( 2)  &     2.154  &      416 &       20 (        4 )\\ 
(1,0)S( 0)  &     2.223  &     2599 &      289 (       58 )\\ 
(2,1)S( 1)  &     2.248  &      521 &       30 (        6 )\\ 
(1,0)Q( 1)  &     2.407  &    10675 &      757 (      454 )\\ 
(1,0)Q( 2)  &     2.413  &     6122 &      616 (      370 )\\ 
(1,0)Q( 3)  &     2.424  &     8936 &      984 (      590 )\\ 
(1,0)Q( 4)  &     2.437  &     4419 &      514 (      308 )\\ 
            \noalign{\smallskip}
            \hline
         \end{array}
     $$ 
   \end{table}

\clearpage

   \begin{table}
      \caption[]{H$_2$ line fluxes measured from the medium-resolution
H- and K-band sum spectra obtained with the HR grism in orders 4 and 3, 
respectively.
      }
         \label{table_HKs}
\centering
     $$ 
         \begin{array}{p{0.25\linewidth}rrrr}
            \hline
            \noalign{\smallskip}
            Line transition      &  \mathrm{Wavelength} & \mathrm{Flux} 
& N\mathrm{(v'J')} \\
&  \mu\mathrm{m} & {10^{-20} \mathrm{W\ m}^{-2}} & 10^{14} \mathrm{cm}^{-2}\\
            \noalign{\smallskip}
            \hline
            \noalign{\smallskip}

(1,0)S( 9)  &     1.688  &      652 &       83 (       17 )\\ 
(1,0)S( 8)  &     1.715  &      278 &       26 (        5 )\\ 
(1,0)S( 7)  &     1.748  &     2328 &      172 (       34 )\\ 
(1,0)S( 6)  &     1.788  &     1504 &       96 (       19 )\\ 
\\
(1,0)S( 2)  &     2.034  &     2977 &      192 (       96 )\\ 
(3,2)S( 5)  &     2.066  &      198 &       11 (        6 )\\ 
(2,1)S( 3)  &     2.074  &      623 &       28 (       14 )\\ 
(1,0)S( 1)  &     2.122  &     8763 &      677 (      135 )\\ 
(2,1)S( 2)  &     2.154  &      371 &       18 (        4 )\\ 
(3,2)S( 3)  &     2.201  &      159 &        8 (        2 )\\ 
(1,0)S( 0)  &     2.223  &     2901 &      323 (       65 )\\ 
(2,1)S( 1)  &     2.248  &      970 &       55 (       11 )\\ 
            \noalign{\smallskip}
            \hline
         \end{array}
     $$ 
%\begin{list}{}{}
%\item[$^{\mathrm{a}}$] This is footnote a
%\end{list}
   \end{table}

   \begin{table}
      \caption[]{Line fluxes measured for atomic transitions using the
low-resolution spectrum obtained with grism GB.}
         \label{table_atoms1}
\centering
$$
         \begin{array}{p{0.25\linewidth}rrr}
            \hline
            \noalign{\smallskip}
            Transition & \mathrm{Wavelength} & \mathrm{Flux}\\
	            & \mu\mathrm{m} & 10^{-20} \mathrm{W\ m^{-2}} \\
            \noalign{\smallskip} \hline \noalign{\smallskip}
    $\mathrm{^1D_2-^3P_2}$ SIII & 0.953 &     839\\ 
    FeII? & 0.955 &     177\\ 
    $\mathrm {^1D_2 - ^3P_1}$ CI   & 0.983 &     179\\ 
    $\mathrm {^1D_2 - ^3P_2}$ CI   & 0.985 &     678\\ 
    H 7--3 & 1.005 &     340\\ 
    ?  & 1.019 &      62\\ 
    $\mathrm {^2P_{3/2} - ^2D_{3/2}}$ SII  & 1.029 &    1019\\ 
    $\mathrm {^2P_{3/2} - ^2D_{5/2}}$ SII  & 1.033 &    2497\\ 
    $\mathrm {^2P_{1/2} - ^2D_{5/2}}$ SII  & 1.038 &     413\\ 
    $\mathrm {^2P_{1/2} - ^2D_{5/2}}$ NI   & 1.041 &    1019\\ 
    HeI  & 1.083 &    5516\\ 
    H 6--3 & 1.094 &     616\\ 
    $\mathrm{^5P - ^5S_0}$ OI   & 1.130 &     195\\ 
    ?    & 1.134 &     192\\ 
    ?    & 1.144 &      59\\ 
    PII  & 1.147 &     116\\ 
    ?    & 1.149 &      39\\ 
    ?    & 1.176 &      58\\ 
$\mathrm{a^6D_{9/2}-a^4D_{5/2}}$  FeII & 1.191 &     143\\ 
$\mathrm{a^6D_{9/2}-a^4D_{7/2}}$  FeII & 1.257 &   11134\\ 
$\mathrm{a^6D_{1/2}-a^4D_{1/2}}$  FeII & 1.271 &     684\\ 
$\mathrm{a^6D_{3/2}-a^4D_{3/2}}$  FeII & 1.279 &    1132\\ 
H 5--3 & 1.282 &    1645\\ 
$\mathrm{a^6D_{5/2}-a^4D_{5/2}}$  FeII & 1.294 &    1650\\ 
$\mathrm{a^6D_{7/2}-a^4D_{7/2}}$  FeII & 1.321 &    3595\\ 
$\mathrm{a^6D_{3/2}-a^4D_{5/2}}$  FeII & 1.328 &    1083\\ 
$\mathrm{a^4F_{9/2}-a^4D_{5/2}}$ FeII & 1.533 &    2439\\ 
$\mathrm{a^4F_{7/2}-a^4D_{3/2}}$ FeII & 1.600 &    1866\\ 

            \noalign{\smallskip}
            \hline
         \end{array}
$$
   \end{table}

   \begin{table}
      \caption[]{Wavelengths and line fluxes of atomic transitions, 
derived from spectra obtained with grisms GR and HR in orders 4 and 3,
respectively}
         \label{table_atoms2}
\centering
     $$ 
         \begin{array}{p{0.25\linewidth}rrr}
            \hline
            \noalign{\smallskip}
            Transition & \mathrm{Wavelength} & \mathrm{flux}\\
	    & \mu\mathrm{m} & 10^{-20} \mathrm{W\ m^{-2}} \\
            \noalign{\smallskip}
            \hline
            \noalign{\smallskip}

$\mathrm{a^4F_{9/2} - a^4D_{5/2}}$ FeII & 1.534 &    2459\\ 
$\mathrm{a^4F_{7/2} - a^4D_{3/2}}$ FeII & 1.600 &    1380\\ 
H 12--4 & 1.641 & 200 \\
$\mathrm{a^4F_{9/2} - a^4D_{7/2}}$ FeII & 1.644 &   13521\\ 
$\mathrm{a^4F_{5/2} - a^4D_{1/2}}$ FeII & 1.664 &     563\\ 
$\mathrm{a^4F_{7/2} - a^4D_{5/2}}$ FeII & 1.677 &    1512\\ 
H 11--4    & 1.681 &     568\\ 
$\mathrm {^3P_0 - ^3D}$ HeI       & 1.700 &     176\\ 
H 10--4    & 1.737 &     972\\ 
H 7--4 &    2.166 &    1191\\ 
\\
$\mathrm{a^4F_{9/2} - a^4D_{5/2}}$ FeII & 1.533 &    2637\\ 
$\mathrm{a^4F_{7/2} - a^4D_{3/2}}$ FeII & 1.600 &    2130\\ 
$\mathrm{a^4F_{9/2} - a^4D_{7/2}}$ FeII & 1.644 &    13830\\ 
$\mathrm{a^4F_{5/2} - a^4D_{1/2}}$ FeII & 1.664 &    1356\\ 
$\mathrm{a^4F_{5/2} - a^4D_{5/2}}$ FeII & 1.677 &    2597\\ 
H 10--4          & 1.736 &     620\\ 
$\mathrm{a^4F_{3/2} - a^4D_{1/2}}$ FeII & 1.745 &    1295\\ 
$\mathrm{a^4F_{3/2} - a^4D_{3/2}}$ FeII & 1.797 &    1017\\ 
$\mathrm{a^4F_{5/2} - a^4D_{5/2}}$ FeII & 1.800 &    1301\\ 
$\mathrm{a^4F_{7/2} - a^4D_{7/2}}$ FeII & 1.809 &    2173\\ 
\\
H 7--4 &   2.166 &     613\\ 
            \noalign{\smallskip}
            \hline
         \end{array}
     $$ 
   \end{table}

\end{document}